\documentclass{optica-article}

\journal{opticajournal} 

\articletype{Research Article}
\usepackage{makecell}
\usepackage{lineno}
\usepackage{multirow}
\usepackage{siunitx}
\usepackage{cite, color, xspace}

\newcommand{\invitro}{\textit{in vitro}\xspace}
\newcommand{\exvivo}{\textit{ex vivo}\xspace}

\newcommand{\exvivoh}{\textit{ex-vivo}\xspace}

\newcommand{\enfaceh}{\textit{en-face}\xspace}

\newcommand{\etal}{\textit{et al.\@}\xspace}
\newcommand{\amean}[2]{{\left\langle{#1} \right\rangle_{#2}\xspace}}
\newcommand{\um}{\(\mu\)m\xspace}

\newcommand{\uM}{\(\mu\)M\xspace}

\newcommand{\nnccs}{{95\% confidence intervals}\xspace}
\newcommand{\dbsq}{{dB$^2$}\xspace}

\graphicspath{{.}{./figures}}

\begin{document}
\title{Neural-network based high-speed volumetric dynamic optical coherence tomography}

\author{%
	Yusong Liu\authormark{1},
	Ibrahim Abd El-Sadek\authormark{1, 2},
	Rion Morishita\authormark{1},
	Shuichi Makita \authormark{1},
	Tomoko Mori\authormark{3},
	Atsuko Furukawa\authormark{3},
	Satoshi Matsusaka\authormark{3} and
	Yoshiaki Yasuno\authormark{1,*}
}

\address{%
	\authormark{1}Computational Optics Group, University of Tsukuba, 1-1-1 Tennodai, Tsukuba, Ibaraki 305–8573, Japan.\\
	\authormark{2}Department of Physics, Faculty of Science, Damietta University, New Damietta City 34517, Damietta, Egypt.\\
	\authormark{3}Clinical Research and Regional Innovation, Faculty of Medicine, University of Tsukuba, Tsukuba, Ibaraki 305-8575, Japan.
}

\email{\authormark{*}yoshiaki.yasuno@cog-labs.org}
\homepage{https://cog-news.blogspot.com/}

\begin{abstract*} 
	We demonstrate deep-learning neural network (NN)-based dynamic optical coherence tomography (DOCT), which generates high-quality logarithmic-intensity-variance (LIV) DOCT images from only four OCT frames.
	The NN model is trained for tumor spheroid samples using a customized loss function: the weighted mean absolute error.
	This loss function enables highly accurate LIV image generation.
	The fidelity of the generated LIV images to the ground truth LIV images generated using 32 OCT frames is examined via subjective image observation and statistical analysis of image-based metrics. 
	Fast volumetric DOCT imaging with an acquisition time of 6.55 s/volume is demonstrated using this NN-based method.
\end{abstract*}

\section{Introduction}
Optical coherence tomography (OCT)\cite{huang1991optical} is a three-dimensional, label-free, and high-resolution imaging modality.
Although standard OCT can visualize the morphological structures of tissue, it cannot assess tissue dynamics such as intracellular motility.
Recently, an OCT-based non-invasive imaging modality called dynamic optical coherence tomography (DOCT) was proposed for visualization and quantification of tissue dynamics\cite{apelian2016dynamic}.
DOCT has been used to investigate \exvivoh tissue metabolism\cite{mukherjee2021label, mukherjee2022label}, the cellular activity of tumor spheroids\cite{ElSadek2020BOE,ElSadek2021BOE,ElSadek2023SR}, and the cellular activity and differentiation processes of alveolar organoids\cite{morishita2023label}, retinal organoids  \cite{Scholler2020Light}, and several other tissues\cite{apelian2016dynamic, ling2017ex, munter2020dynamic, park2021quantitative, scholler2019probing, thouvenin2017cell, thouvenin2017dynamic}.

DOCT is a method that combines time-sequential OCT acquisition at the same location with analysis of the temporal fluctuations of the OCT signal.
Several signal processing methods have been proposed to perform the temporal fluctuation analysis, including signal fluctuation-magnitude analysis\cite{ElSadek2020BOE, ElSadek2021BOE}, fluctuation speed analysis\cite{ElSadek2020BOE, ElSadek2021BOE}, time-frequency analyses\cite{Scholler2020Light, leung2020imaging, mclean2017frequency}, and eigen-decomposition variance analysis\cite{wei2019dynamic}.

Although DOCT has been used successfully to assess several dynamic processes of various tissues, the process requires the acquisition of large numbers (i.e., from tens to thousands) of repeated OCT frames within a long time separation from the first to the last frame at a single location in the sample, which then leads to long total acquisition time.
Specifically, a frequency-constrained robust principal component analysis demonstrated by McLean \etal required 1,350 frames per location\cite{mclean2017frequency}, while a time-frequency-analysis based DOCT demonstrated by Leung \etal required 1,000 frames per location\cite{leung2020imaging}.
Although our developed  DOCT methods, the logarithmic intensity variance (LIV) and OCT correlation decay speed (OCDS) methods, require relatively small numbers of frames, e.g., 16 or 32 frames, the volumetric measurement time is still long and ranges from approximately 30 s up to 1 min \cite{ElSadek2021BOE}.
The long acquisition time hampers high-throughput imaging, which is important for certain applications, e.g., large-sample-number drug screening.

Recently, use of deep-learning neural networks (NNs)\cite{o2015introduction, schmidhuber2015deep} to generate high-quality OCT angiography (OCTA) images from small numbers of OCT frames has been demonstrated\cite{jiang2020comparative}.
Because both OCTA and DOCT involve the analysis of time-sequential OCT images, we hypothesized that NNs can also generate high-quality DOCT images from small numbers of OCT frames, and such an approach will enable high-speed volumetric DOCT imaging.

In this study, we demonstrate deep-learning NN-based DOCT generation, in which an NN is trained to generate a high-quality LIV image from only four OCT frames.
The LIV is one type of DOCT and is defined as the time variance of the logarithmic OCT signals.
The NN-generated LIV image was compared subjectively and objectively with conventional LIV images.
The comparison showed high consistency between the NN-generated LIV and conventional LIV images.
It is also demonstrated that a high-quality LIV volume of a tumor spheroid can be obtained from a volumetric OCT dataset that was acquired within only 6.55 s, while a conventional LIV measurement of the same volume size requires an acquisition time of 52.4 s.

\section{Principle}\label{sec:Principle}

\subsection{Neural network model}\label{sec:NNmodel}
\begin{figure}
	\centering\includegraphics[width=13cm]{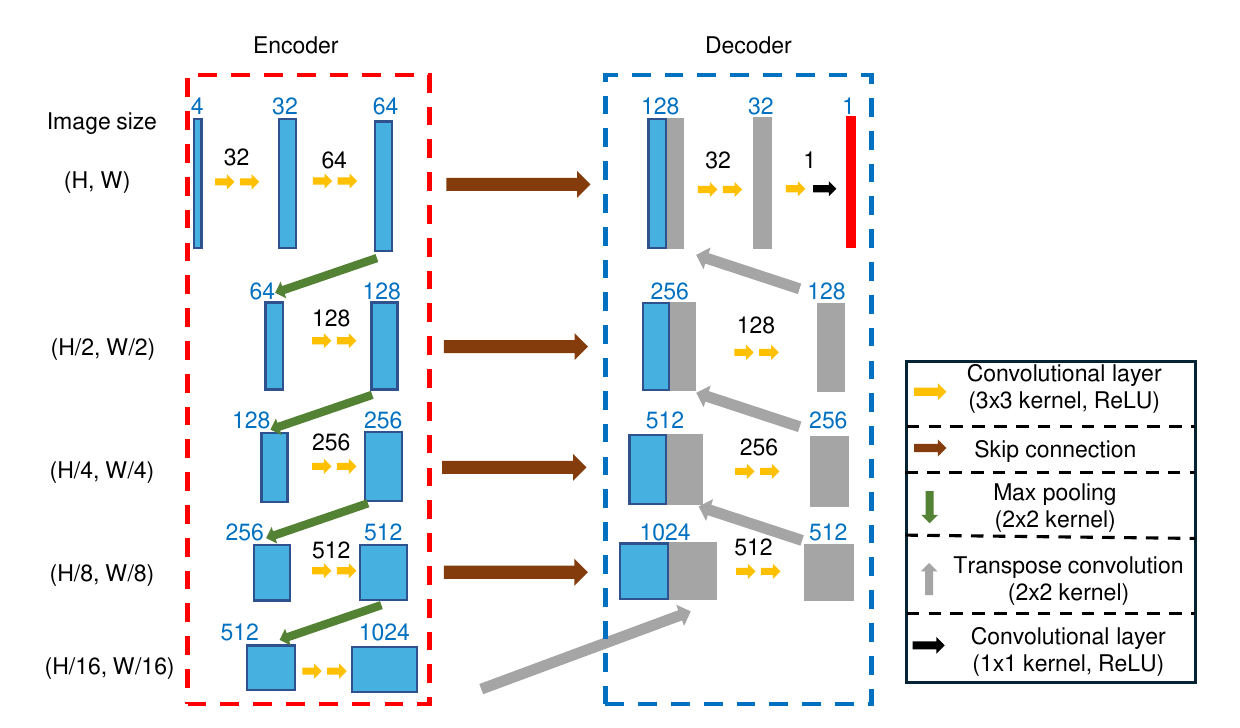}
	\caption{%
		NN architecture used for LIV generation. 
		The NN consists of three parts: an encoder (red dashed-line box), a decoder (blue dashed-line box), and a skip connection (brown arrows). 
		The input to the NN is a set of four cross-sectional OCT images, which were acquired at the same position in the sample but at different time points,  and the output is an LIV image, which is a DOCT image that is defined as the time variance of the dB-scaled OCT images. 
		The leftmost part indicates the spatial dimensions of the image at each step of the U-Net, where (H, W) represents the original size of the input image.
	}
	\label{fig:NNArchitecture}
\end{figure}

The architecture of the NN that generates the high quality LIV image is shown in Fig.\@ \ref{fig:NNArchitecture}.
This architecture is a modified version of the U-Net architecture \cite{ronneberger2015u}.
The main structure in this network is its convolutional layer, which consists of a convolution operation and a rectified-linear-unit (ReLU) activation function.
The input has four channels that correspond to the four time-sequential cross-sectional logarithmic OCT images, and the output is an LIV image.
Convolutional layers are used to extract the image features and to change the number of channels.
The black numbers over the convolutional layers (yellow and black arrows) shown in Fig.\@ \ref{fig:NNArchitecture} indicate the numbers of kernels of convolution.

The NN comprises three parts: the encoder (red dashed-line box), the decoder (blue dashed-line box), and the skip connections (the brown arrows between the encoder and decoder).
The encoder extracts features of multiple spatial scales, while the decoder reconstructs the output image using these spatial features. 
Skip connections pass the multi-scale features from the encoder to the decoder, thereby preserving high-resolution information.
The input (i.e., the set of four OCT images) first passes through the encoder.
The encoder is realized using 12 convolutional layers with a 3$\times$3 kernel (yellow arrows) and four max-pooling operations with a 2$\times$2 kernel (green arrows).
Each max-pooling operation reduces the two-dimensional image size by a factor of four, i.e., by a factor of two for each dimension.
After the encoder, the image size is reduced by a factor of 16 for each dimension, and the number of image channels is increased from 4 to 1,024.

The decoder consists of combinations of transpose convolutional layers (gray arrows) and convolutional layers (yellow arrows and one black arrow). 
Each transpose convolutional layer uses a 2$\times$2 kernel and enlarges the image by a factor of two for each dimension.
The output from the convolutional layer is concatenated with an intermediate output from the encoder through a skip connection (brown arrow).
The concatenation operation via the skip connection allows feature sharing between the encoder and the decoder.
The concatenated image then passes through two convolutional layers, where each layer uses a 3$\times$3 kernel.
After four sets composed of a transpose convolutional layer and two convolutional layers, and four skip connection operations, the image size becomes the same as that of the original input image, and the number of channels is reduced from 1,024 to 32.
Finally, a set composed of a convolutional layer with a 3$\times$3 kernel and a subsequent second convolutional layer with a 1$\times$ 1 kernel (black arrow) reduces the number of channels from 32 to 1.

\subsection{Target image and input dateset}
\label{sec:TargetAndInput}
The NN model is trained to generate a ground truth LIV image from a set of four time-sequential dB-scaled OCT images.
The ground truth LIV image is defined as the time variance of a dB-scaled OCT image that is computed from 32 time-sequential OCT frames \cite{ElSadek2020BOE} as follows: 
\begin{equation}
	\mathrm{LIV}(x, z) = \amean{\left(I_{dB}(x, z, t_i)-{\amean{I_{dB}(x, z, t_i)}{i}}\right)^2}{i},
	\label{eq:defLiv}
\end{equation} 
where $I_{dB}(x, z, t_i)$ represents a dB-scaled OCT intensity image at the spatial position $(x, z)$, and $x$ and $z$ are the lateral and depth positions, respectively. 
$t_i$ is the time of acquisition of the $i$-th frame.
$\amean{\quad}{i}$ represents the average over time.

In our typical implementation, the number of frames is 32 and the time separation between consecutive frames denoted by $t_{i+1} - t_{i}$ is 204.8 ms, which means that the entire time sequence is acquired in 6.35 s.
The implementation of this time-sequential data acquisition process is described later in more detail in Section \ref{sec:32protocol}.

The four input frames are also dB-scaled OCT images and are acquired with a time separation of 1,638.4 ms, i.e., with an eight-fold longer time separation when compared with that of the ground truth LIV image.
The total time separation from the first frame to the last is 4.92 s.
The four dB-scaled OCT frames are then concatenated into a four-channel dataset to be fed into the NN.

\subsection{Training flow}
\label{sec:Training flow}
\begin{figure}
	\centering\includegraphics[width=13cm]{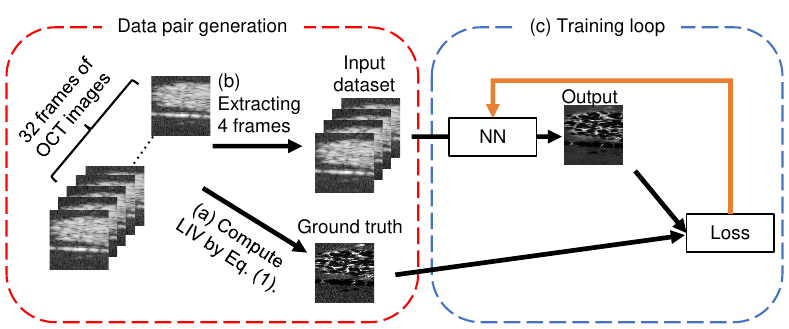}
	\caption{%
		Schematic diagram of the data pair generation process for NN training and the training flow.
		The original data are a time-sequence of 32 OCT images (frames) with an inter-frame time separation of 204.8 ms.
		(a) The ground truth LIV image is computed as the pixel-wise time variance among the 32 OCT images.
		(b) Four frames with an inter-frame time separation of 1,638.4 ms are then extracted from the 32-frame OCT sequence, and these frames are paired with the ground truth LIV image.
		(c) In the training loop, the NN accepts the four-frame sequence as an input, and the network parameters are then updated to generate the ground truth LIV image.
		The loss is computed by a customized loss function (the weighted mean absolute error; see Section \ref{sec:Training details}) from the network output and the ground truth, and this loss is then back-propagated to the NN model to update the parameters (as indicated by the orange arrow).
	}
	\label{Fig:TraingFlow}
\end{figure}

For the training of the NN model, we use time-sequential OCT frames to generate both the ground truth and the input set, as depicted schematically in Fig.\@ \ref{Fig:TraingFlow}.
In this particular study, the original OCT time-sequence consists of 32 frames and the ground truth is obtained from all 32 frames using Eq.\@  (\ref{eq:defLiv}) [part (a) of the figure].
The input time sequence is then constructed by extracting the 8th, 16th, 24th, and 32nd frames from the original dataset [part (b)].
In the training process, the loss is computed based on the output from the NN model and the ground truth LIV, and is then back-propagated to the NN to update the parameters [part (c)].
A detailed description of the implementation of the training process, including the definition of the loss function and selection of the hyper parameters, is presented later in Section \ref{sec:Training details}.

\section{Implementation}
\label{sec:implementation}
\subsection{OCT device}
\label{sec:JMOCT}
A custom-built Jones-matrix OCT (JM-OCT) device is used for data acquisition. 
This system is identical that was used in our previous LIV-imaging studies of both \invitro samples \cite{ElSadek2020BOE, ElSadek2021BOE, ElSadek2023SR, morishita2023label} and \exvivo samples \cite{mukherjee2021label, mukherjee2022label}.
Because full details of this system can be found elsewhere \cite{li2017three, Miyazawa2019BOE}, we only describe the system specifications briefly here.
The system comprises a swept-source OCT with a microelectromechanical systems (MEMS)- based scanning light source (AXP50124-8, Axsun Technologies, MA).
The center wavelength of the probe beam is 1.3 \um and the beam scanning speed is 50 kHz.
The full-width-at-half-maximum axial resolution is 14 \um in tissue and 19 \um in the air.
The 1/{e$^2$}-width lateral resolution is 19 \um.

Although the JM-OCT is a polarization-sensitive OCT and provides four OCT images corresponding to the four polarization channels, we use only a polarization-insensitive OCT intensity image that represents the intensity average of the four OCT images.
In other words, our NN-based method does not use polarization-sensitive information and is thus compatible with conventional polarization-insensitive OCT devices.

\subsection{Volumetric scan protocol for 32-frame sequence}
\label{sec:32protocol}
\begin{figure}
	\centering\includegraphics[width=13cm]{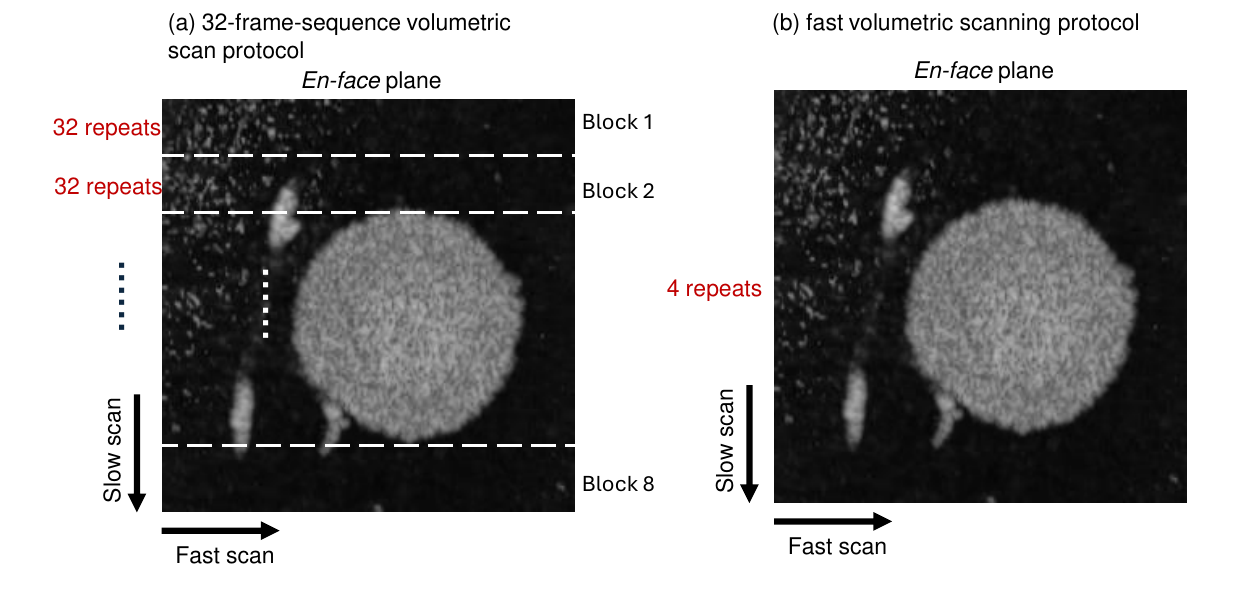}
	\caption{
		Schematic diagram of the two scanning protocols used for volumetric data acquisition.
		(a) A 32-frame-sequence volumetric scan protocol is used to acquire a dataset from which both the ground truth LIV images and the NN-based LIV images are generated.
		In this protocol, the entire lateral field of view is split into eight blocks, and each block is scanned repeatedly using a raster scan protocol with 16 B-scan locations for a total of 32 times.
		Therefore, for the eight blocks, a total of 128 B-scan locations are scanned.
		The inter-frame time separation at each B-scan location is 204.8 ms and the total data acquisition time is 52.4 s/volume.
		(b) A fast volumetric scanning protocol is used to demonstrate fast volumetric LIV imaging based on the NN-based LIV method. 
		The complete field is scanned repeatedly using a raster scan protocol with 128 B-scan locations for a total of four times.
		The inter-frame time separation at each B-scan location is 1,638.4 ms and the total data acquisition time is 6.55 s/volume.
	}
	\label{fig:ScanningProtocol}
\end{figure}
For the volumetric images, we acquire 32 frames at each of 128 B-scan locations.
To obtain a time separation between frames of 204.8 ms and keep the total volumetric acquisition time short  (e.g., less than 1 min), we used the repeating raster scan method described in Ref.\@ \cite{ElSadek2021BOE}.

As shown in Fig.\@ \ref{fig:ScanningProtocol}(a), the \enfaceh field is split into eight blocks along the slow scan direction.
Each block is then scanned 32 times using the volumetric raster scan protocol.
In this case, each cross-section (i.e., each frame) consists of 512 A-lines, 32 frames are acquired at each B-scan location, and one block consists of 16 B-scan locations.
After this repeating raster scan is performed sequentially for the eight blocks, a volumetric time sequential OCT dataset is obtained.
To form the final volumetric dataset, 128 B-scan locations were scanned.
The inter-frame time interval was 204.8 ms and the time separation between the first and last frames at each location was 6.35 s.
The total acquisition time for the volume image was 52.4 s. 
The datasets required to train the NN model were acquired using this protocol.

\subsection{Samples for NN training and evaluation}
\label{sec:Sample}
To train and evaluate the NN, 60 breast-cancer spheroids (MCF-7 cell line) and 60 colon cancer spheroids (HT-29 cell line) were used.
These spheroids were cultivated using 96-well plates, and each spheroid was seeded with 1,000 cells.
To increase the variety of these spheroids, the spheroids were treated with anti-cancer drugs, including paclitaxel (PTX; for MCF-7) and SN-38 (for HT-29) with concentrations of 0, 0.1, 1, and 10 \uM, where 0 \uM means that the spheroid was not treated using the drug.
The treatment times were 1, 3, or 6 days, which are denoted by Day-1, -3, and -6, respectively, throughout the manuscript.
Five spheroids were prepared for each combination of drug concentration and treatment time.
Note that the spheroids and their OCT data are identical to those used in our previous drug-response investigation study.
More details of the samples and their cultivation protocols can therefore be found in our previous publication \cite{ElSadek2023SR}.

Each of the spheroids was scanned with a lateral field of view (FOV) of 1 mm $\times$ 1 mm.
The five spheroids for each condition were split into groups of three, one, and one for training, validation, and evaluation, respectively.
As a result, in the NN training process, we used 96 spheroids (i.e., 72 for training and 24 for validation).
The remaining 24 spheroids were reserved for the evaluation study, which is described in detail in Section \ref{sec:evaluationmethod}.

\subsection{Implementation neural network}
\label{sec:Implementation for NN training}

\subsubsection{Data sets for training}
\label{sec:data setForTraining}
To generate the training and validation datasets, the 72 (training) and 24 (validation) spheroids were scanned using the 32-frame-sequence volumetric scan protocol that was described in Section \ref{sec:32protocol}.
From each volume, the ground truth LIV and a corresponding input four-frame sequence were generated in the manner described in Section \ref{sec:TargetAndInput}.

From each volumetric dataset, we selected 20 B-scan locations, and the cross-section at each selected location contained a sample region of at least 1,000 pixels.
Then, from each cross-section, we extracted 40 patch pairs with 64 $\times$ 64 pixels, such that 800 image patches were extracted from each spheroid, where the term ``patch pair'' means a pair composed of the input and the ground truth.
Therefore, for the training and validation processes, the sizes of both the input image and the ground truth LIV image are 64 $\times$ 64 pixels.
The extraction location for each patch is selected at random, but all the patches contain the spheroid region at least partially.
The patch locations can also be partially overlapped.
The final training and validation datasets consist of 57,600 and 19,200 patch pairs, respectively.

\subsubsection{Detailed implementation of neural network and its training}
\label{sec:Training details}
The NN was implemented in Python 3.7 using the open source machine learning platform TensorFlow 2.6 on a PC equipped with a graphics processing unit (GPU; NVIDIA GeForce RTX 3090 with 10,496 Compute Unified Device Architecture (CUDA) cores and 24 GB of memory).
The NN model was trained based on mini batches, where the total of 57,600 patch pairs in the training dataset was divided into 800 mini batches.
Each mini batch consisted of 72 image patch pairs, and all patch pairs were taken from 72 different spheroids.

The ground truth LIV image is not evenly distributed in terms of its value (as will be discussed in Section \ref{sec:lossfunctionselection}), and thus we used the weighted mean absolute error (wMAE) as a loss function.
In this case, we customized the weight (W) to increase the weights of the high-LIV regions as follows
\begin{equation}
	\mathrm{W}(x, z, b; \theta) = 
	\begin{cases}
		2 & \text{for} \quad \mathrm{LIV}(x, z; b) \geq \theta\\
		1 & \text{otherwise}
	\end{cases},
	\label{Eq: weight}
\end{equation}
where $\mathrm{LIV}$ is the ground truth LIV image defined using Eq.\@ (\ref{eq:defLiv}), $(x, z)$ is a spatial position within the batch, $b$ is the batch index, and $\theta$ is a predefined threshold for the LIV.
As shown in the equation above, we set higher weights for the pixels whose ground truth LIV values are larger than the threshold value ($\theta$).
In this particular study, we set $\theta$ empirically at 9 dB$^{2}$.

The wMAE was then computed from the NN output and the ground truth using this weight as follows
\begin{equation}
	\mathrm{wMAE}(\theta) =\frac{\sum_{x, z, b}{\mathrm{W}(x, z, b; \theta) \circ {\left| \mathrm{LIV}(x, z; b) - \mathrm{LIV'}(x, z; b)\right|}}}
	{\sum_{x, z, b} \mathrm{W}(x, z, b; \theta)}.
	\label{Eq: wMAE}
\end{equation}
where $\mathrm{LIV}$ and $\mathrm{LIV'}$ represent the ground truth LIV and the NN output, respectively, and $\circ$ represents the element-wise product.
The rationality of this loss function and the optimal threshold selection procedure will be discussed later in Section \ref{sec:lossfunctionselection}.

The NN model parameters were updated using the Adam optimizer\cite{kingma2014adam} with the wMAE loss.
The Adam optimizer is a first-order gradient-based optimizer that incorporates adaptive momentum estimates. 
Due to its robustness, it is widely used in deep learning and is also suitable to train our NN model. 
To enable the NN model to learn the detailed image pattern, we used a decaying learning rate strategy\cite{you2019does}, in which the learning rate is defined as $10^{-4} + 5 \times 10^{-4} / \mathrm{epoch}$.
To prevent over-fitting, the training process was stopped when the validation loss did not decrease for seven consecutive epochs, and the parameters of the eighth epoch from the last epoch were stored as the trained NN model.

The training takes roughly 19 minutes with the PC described in the first paragraph of this section.
The inference time (i.e., image generation time) of a volumetric LIV with the dimensions of 128 $\times$ 384 $\times$ 512 pixels is approximately 15.82 s.

\section{Protocol and method of performance evaluation study}
\label{sec:evaluationmethod}
\subsection{Image types}
We used three image types in the performance evaluation of our NN-based method.
The first image type was a standard LIV image computed from the 32 dB-scaled OCT frames, and this LIV image was identical to the ground truth image.
This type is designated C32LIV, where C stands for ``conventional.''
The second image type was the output of the NN model, i.e., the image generated by using our proposed method.
This image type was computed from four dB-scaled OCT frames, and is designated NN4LIV.
The last image type is the time variance of the four dB-scaled OCT images which were identical to those used for the NN4LIV image.
Because the method of computation used for this type of image is identical to that used for the conventional LIV image [Eq.\@ (\ref{eq:defLiv})] and the number of images used is four, this image type is designated C4LIV.
This image is used as a reference for comparison with C32LIV and NN4LIV.

To enable subjective observation of the image, pseudo-color images were created for all image types, where the image brightness is the OCT intensity and the color (hue) of the image is one of C32LIV, NN4LIV, or C4LIV.
The details of the color image formation process can be found in Section 2.3 of Ref.\@ \cite{ElSadek2020BOE}.

\subsection{Samples for evaluation study}
\label{sec:evaluationSamples}
We used 12 breast cancer spheroids and 12 colon cancer spheroids to evaluate the trained NN model.
It should be noted that only one NN model was trained for both types of spheroids.
To assess the repeatability, each spheroid was scanned twice consecutively using the 32-frame-sequence volumetric scan protocol.
These two consecutive measurement sessions are named S1 and S2.
The separation between the starting times for S1 and S2 was approximately 2 min.
The three image types, i.e., C32LIV, NN4LIV, and C4LIV, were computed for each volumetric dataset.

Note that each OCT cross-section of these spheroids consists of 384 depth pixels $\times$ 512 transverse pixels.
Unlike the NN training case, the full-size image is fed into the trained NN model at once, thus allowing the full-size cross-sectional LIV image to be obtained via a single inference operation.

\subsection{Image evaluation metrics}
\label{sec:Image evaluation metrics}
In this study, we used several image-based evaluation metrics that had also been used to quantify spheroid viability in our previous spheroid-based drug response studies \cite{ElSadek2021BOE, ElSadek2023SR}.
These evaluation metrics include the mean LIV and the viable cell ratio (VCR) within the spheroid region.
To compute these metrics, we first segmented the spheroid region using a semi-automatic OCT intensity threshold-based segmentation method\cite{ElSadek2021BOE}.
After excluding the B-scans that did not contain the spheroid, 1,312 cross-sectional segmentation masks for 24 volumes were obtained.
The segmentation was performed using the S1 dataset, and the same segmentation mask was also used to process the S2 dataset.
To apply the mask that was computed from the S1 data to the S2 datasets, small mutual axial shifts between the S1 and S2 datasets were computed and 
were then corrected. 

The VCR is defined as the ratio of the number of pixels with high LIV values to the total number of pixels within a spheroid, where the high-LIV pixels were defined using a predefined LIV threshold.
This threshold was defined empirically as 3 dB$^2$\cite{ElSadek2020BOE}.
The VCR was computed for each cross-sectional LIV image and was also computed for the entire three-dimensional volume of the spheroid.
Here, we call them the ``VCR of each B-scan,'' and the ``VCR of each volume,'' respectively.

In addition, the mean LIVs were calculated for small regions of interest (ROIs)  and for the complete spheroid.
The mean LIVs of the entire spheroid region were computed for both the complete spheroid volume and for each B-scan.
Here, we call these them the ``mean LIV of each volume'' and the ``mean LIV of each B-scan,'' respectively.

For the mean LIV of the ROIs, we selected two small ROIs manually, where one was located at the spheroid core and the other was located at the spheroid periphery; these ROIs are similar to those used in Ref.\@ \cite{ElSadek2020BOE}.
The spheroid core typically shows low LIV because of necrosis, whereas the spheroid periphery shows high LIV because of its high viability.
To perform this ROI-based analysis, we only used spheroids that showed clear core and periphery appearances in the LIV images.
Five spheroids were selected from the total of 24 spheroids, and all five were MCF-7 spheroids.
We then manually selected 10 cross-sectional images from each selected volume.
All the selected cross-sections showed clear core-to-periphery contrast in the LIV images.
Finally, one ROI was selected for each of the core and the periphery for each cross-section, which meant that we selected a total of 10 core ROIs and 10 periphery ROIs for each spheroid.
The physical size of each ROI is 20 (depth) $\times$ 60 (transverse) pixels, which corresponds to 114 \um (depth) $\times$ 117 \um (transverse).
The ROI regions selected using dataset S1 were also applied to the S2 dataset after the shift correction procedure.
All ROI selections were based on the C32LIV image and the same ROIs were also used for the NN4LIV and C4LIV images.
Finally, the mean LIV was computed for each ROI at each cross-section, and the results were then used to perform data visualization (Fig.\@ \ref{plot:agreementMetrics}).
Additionally, the mean values of all core ROIs and all periphery ROIs were computed for each spheroid, and the results were used to perform the intraclass correlation analysis that will be described in the next section (Section \ref{sec:Statistical analysis}).
Note that we did not use the image metrics (means of the LIV and VCR) from each B-scan to perform the correlation analysis because these metrics, when acquired from the same spheroid, are mutually correlated in principle.

\subsection{Statistical analysis}
\label{sec:Statistical analysis}
The agreements in terms of the evaluation metrics among the different image types (i.e., NN4LIV and C4LIV versus C32LIV) of S1 were evaluated using the intraclass correlation coefficient (ICC) based on a single rating, absolute agreement, two-way mixed-effects model implemented using the Pingouin package (ver. 0.5.3) \cite{vallat2018pingouin} on Python 3.7.
The repeatability  was also evaluated by computing the same ICC between datasets S1 and S2 for each metric and for each image type (i.e., for C32LIV, NN4LIV, and C4LIV).
The criteria for the ICC interpretation are given as follows.
ICC values of less than 0.5, between 0.5 and 0.75, between 0.75 and 0.9, and more than 0.9 represent poor, moderate, good, and excellent agreement (or repeatability), respectively\cite{koo2016guideline}.

\subsection{Demonstration of fast-volumetric measurement}
\label{sec:Four-frame scanning protocol}
Thus far, we have described methods based on the datasets obtained when using the 32-frame-sequence volumetric scan protocol (as described in Section \ref{sec:32protocol}), which has a volumetric acquisition time of 52.4 s/volume.
In the studies described above, at each B-scan location, the NN4LIV image was generated using four frames extracted from a 32-frame sequence, and the volumetric data acquisition of the NN4LIV image took 52.4 s.
In practical use of the NN4LIV image, we do not need to acquire 32 frames at a single location, and four frames are sufficient.
As a result, the volumetric acquisition process can be faster.

To demonstrate the NN4LIV image obtained with this fast volumetric acquisition, we designed another fast volumetric scanning protocol that gives four frames at each B-scan location, and these four frames have the same inter-frame separation time as the four frames extracted from the frame sequence obtained by the 32-frame-sequence protocol.
This fast acquisition protocol is a simple repeating raster scan and is depicted in Fig.\@ \ref{fig:ScanningProtocol}(b).
Each raster scan consists of 128 B-scan locations and each frame consists of 512 A-lines.
And the raster scan is repeated four times.
Therefore, the inter-frame time separation of these four frames is 1638.4 ms, and the time window (i.e., the time separation from the first frame to the last frame at each B-scan location) is 4.92 s. 
These times are identical to those of the extracted four frames of the 32-frame-sequence volumetric scan protocol.
But the total volumetric acquisition time is only 6.55 s/volume, which is 12.5\% of the corresponding time for the 32-frame-sequence volumetric scan protocol.

It should be noted that if we scan only a single cross-sectional position, the reduction in acquisition time is not significant. 
Assuming $\Delta t$ represents the inter-frame time separation of a 32-frame acquisition, the inter-frame acquisition time of the four frames used to feed the NN is 8$\Delta t$. 
Therefore, the total acquisition time of the four frames is (4-1) $\times$ 8$\Delta t$ = 24$\Delta t$, whereas that of the 32-frame sequence is (32-1)$\times\Delta t$ = 31$\Delta t$. Consequently, the total acquisition time of a single cross-section with the four-frame scan is 77.4\% of the 32-frame scan, resulting in only a 22.6\% reduction in acquisition time. 
However, since the majority of applications of DOCT require volumetric imaging, this limited reduction in cross-sectional imaging time is not a significant limitation of our NN method.

\section{Results}
\subsection{Performance evaluation study results}
\label{sec:NN evaluation results}

\subsubsection{Image observations}
\label{observationresults}
\begin{figure}
	\centering\includegraphics[width=13cm]{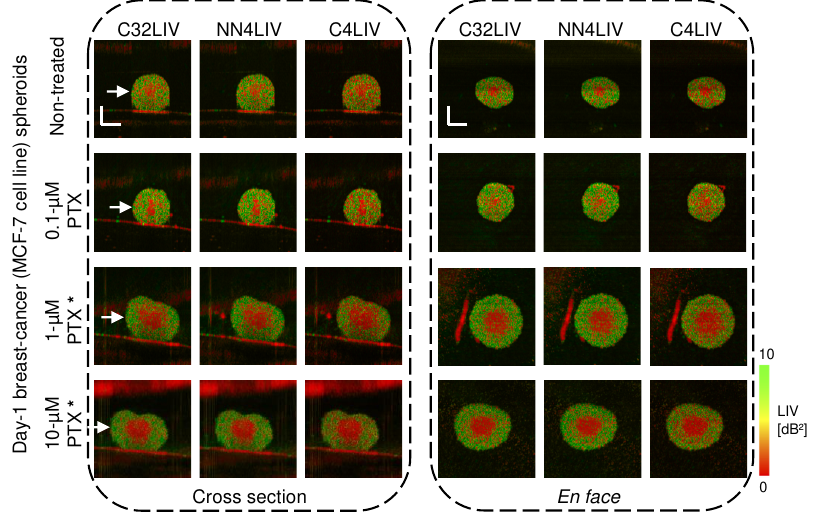} 
	\centering\includegraphics[width=13cm]{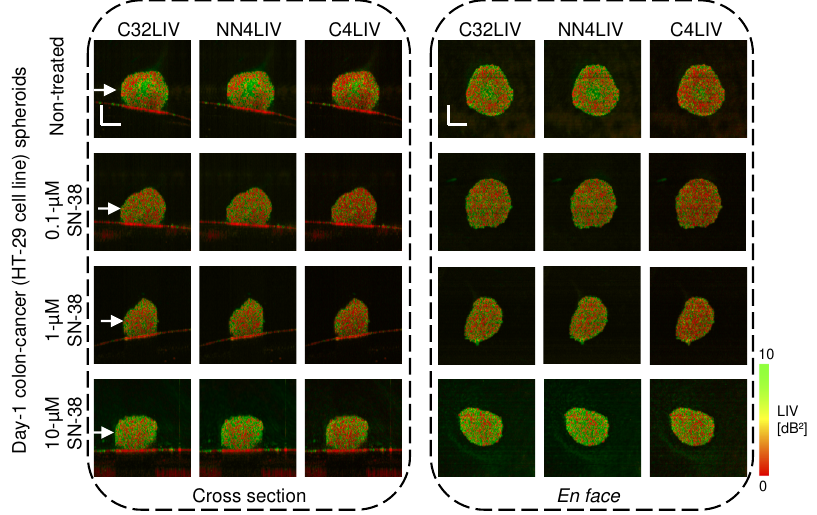}
	\caption{%
		Comparison of the C32LIV, NN4LIV, and C4LIV images obtained from the same 32-frame-sequence OCT images.
		In the images, ``*'' indicates the samples that were used for selection of the core and periphery ROIs, which are used later for the statistical analysis.
		The upper and lower halves of the figure show pseudo-color images of the breast cancer (MCF-7) spheroids and the colon cancer (HT-29) spheroids, respectively.
		These breast and colon cancer spheroids were treated using PTX and SN-38, respectively.
		The left and right halves of the figure show the cross-sectional and \enfaceh images, which were both obtained at the approximate centers of the spheroids.
		The scale bar represents 200 \um.
		NN4LIV images reproduce the spatial patterns and the values of the ground truth C32LIV image well, whereas the C4LIV images show more low-LIV pixels (the red pixels at the periphery) than the C32LIV images.}
	\label{Fig:evaluationImage}
\end{figure}

Figure \ref{Fig:evaluationImage} shows the cross-sectional and \enfaceh C32LIV, NN4LIV, and C4LIV images of the breast cancer (MCF-7) spheroids (upper half of the figure) and the colon cancer (HT-29) spheroids (lower half) from treatment Day-1.
The ``*'' beside an image in Fig.\@ \ref{Fig:evaluationImage} indicates a sample that was used in the selection of the core and periphery ROIs.
All images were acquired from the first measurement session (S1).
The LIV images of all the other spheroids and the second measurement session (S2) can be found in the figures in Supplement 1 (Figs.\@ S1 to S10).
For both spheroid types, we can see that the NN4LIV and C32LIV images show similar spatial patterns and values (i.e., similar colors in the pseudo-color images).
The C4LIV images show more low-LIV signal pixels (i.e., red pixels), which make the necrotic core and the vital periphery less distinguishable.
These results suggest that the appearance of each NN4LIV image is reproduced well and is consistent with the appearance of the corresponding C32LIV image, whereas the C4LIV images are not really consistent with the C32LIV images.

\subsubsection{Agreements of NN4LIV and C4LIV versus C32LIV}
\label{AgreementResult}
\begin{figure}
	\centering\includegraphics[width=13cm]{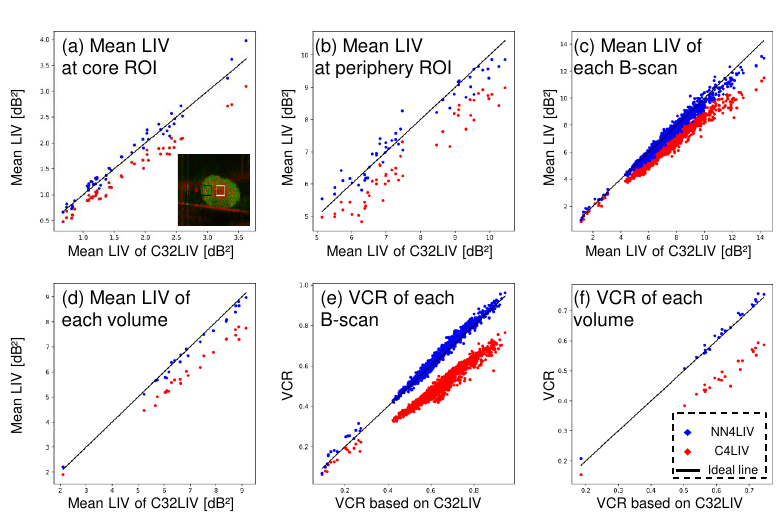}
	\caption{%
		Agreement evaluation of the image matrices obtained from NN4LIV and C4LIV images against those obtained from C32LIV images.
		The mean LIVs at (a) core ROI, (b) periphery ROI, (c) of each B-scan and (d) of each volume, and the VCRs of (e) each B-scan and (f) each volume of the NN4LIV (blue) and C4LIV (red) images are plotted versus those of the C32LIV (ground truth) images.
		Black lines indicate perfect agreement.
		The metrics of the NN4LIV images show high agreement with those of the C32LIV images, while the metrics of the C4LIV images clearly show lower values than those of the C32LIV images.
		The inset in (a) shows the example of the core ROI and the periphery ROI (white and black boxes, respectively) that were used to compute the metrics.}
	\label{plot:agreementMetrics}
\end{figure}
The image metrics, i.e., the mean LIVs at (a) the core ROI, (b) the periphery ROI, (c) of each B-scan and (d) of each volume, and the VCRs of (e) each B-scan and (f) each volume, are plotted versus the corresponding metrics of the ground truth C32LIV images in Fig.\@ \ref{plot:agreementMetrics}.
The blue and red spots represent the NN4LIV and C4LIV data, respectively, and the black lines represent the perfect agreement with the C32LIV data that denotes an ``ideal line.''
For all six metrics, the NN4LIV metrics are close to the ideal line, i.e., the NN4LIV metrics are close to the corresponding ground truth C32LIV metrics.
In contrast, the C4LIV results evidently show lower metric values than the ground truth data.

The agreements of the NN4LIV and C4LIV data with the C32LIV were evaluated quantitatively by computing the ICCs of four of the six metrics.
Intraclass correlation of the mean LIVs and VCRs of each B-scan were not examined here because the B-scans (i.e., the cross-sectional LIV images) of the same spheroids cannot be independent of each other. 
The ICC results and their 95\% confidence intervals are summarized in Table \ref{tab:AgreementMain}.
The NN4LIV metrics show ``excellent'' agreement with the C32LIV results for all four metrics, i.e., ICC $>$ 0.9.
In contrast, the C4LIV metrics show only ``good'' (0.75 $\leq$ ICC $<$ 0.9 for mean LIVs) or ``poor'' (ICC $<$ 0.5 for VCR) agreements.
In addition, all the C4LIV metrics show very wide 95\% confidence intervals and their lower bounds are very low (as indicated by red numbers).
Therefore, C4LIV cannot provide a reliable alternative to C32LIV.

\begin{table}[]
	\caption{%
		ICCs and their 95\% confidence intervals (written in [ ])  for the NN4LIV and C4LIV metrics versus those of C32LIV (the ground truth).
		Higher (closer to 1.0) ICC values indicate better agreement.
		All NN4LIV metrics show excellent agreement with the C32LIV metrics.
		In contrast, the C4LIV metrics show only poor to good agreement levels, and the lower bounds of their 95\% confidence intervals are very low. 
	}
	\centering
	\begin{tabular}{c|c|c}
		& NN4LIV vs C32LIV & C4LIV vs C32LIV \\
		\hline
		Mean LIV at core ROI & 0.997 [0.96,1.0] & 0.880 [\textcolor{red}{-0.04}, 0.99] \\ 
		\hline
		Mean LIV at periphery ROI & 0.988 [0.90, 1.0] & 0.819 [\textcolor{red}{-0.03}, 0.98] \\ 
		\hline
		Mean LIV of each volume & 0.991 [0.92, 1.0] & 0.808 [\textcolor{red}{-0.03}, 0.96] \\ 
		\hline
		VCR of each  volume & 0.986 [0.96, 0.99] &0.487 [\textcolor{red}{-0.02}, 0.82] \\ 
	\end{tabular}
	\label{tab:AgreementMain}
\end{table}

\begin{figure}
	\centering\includegraphics[width=13cm]{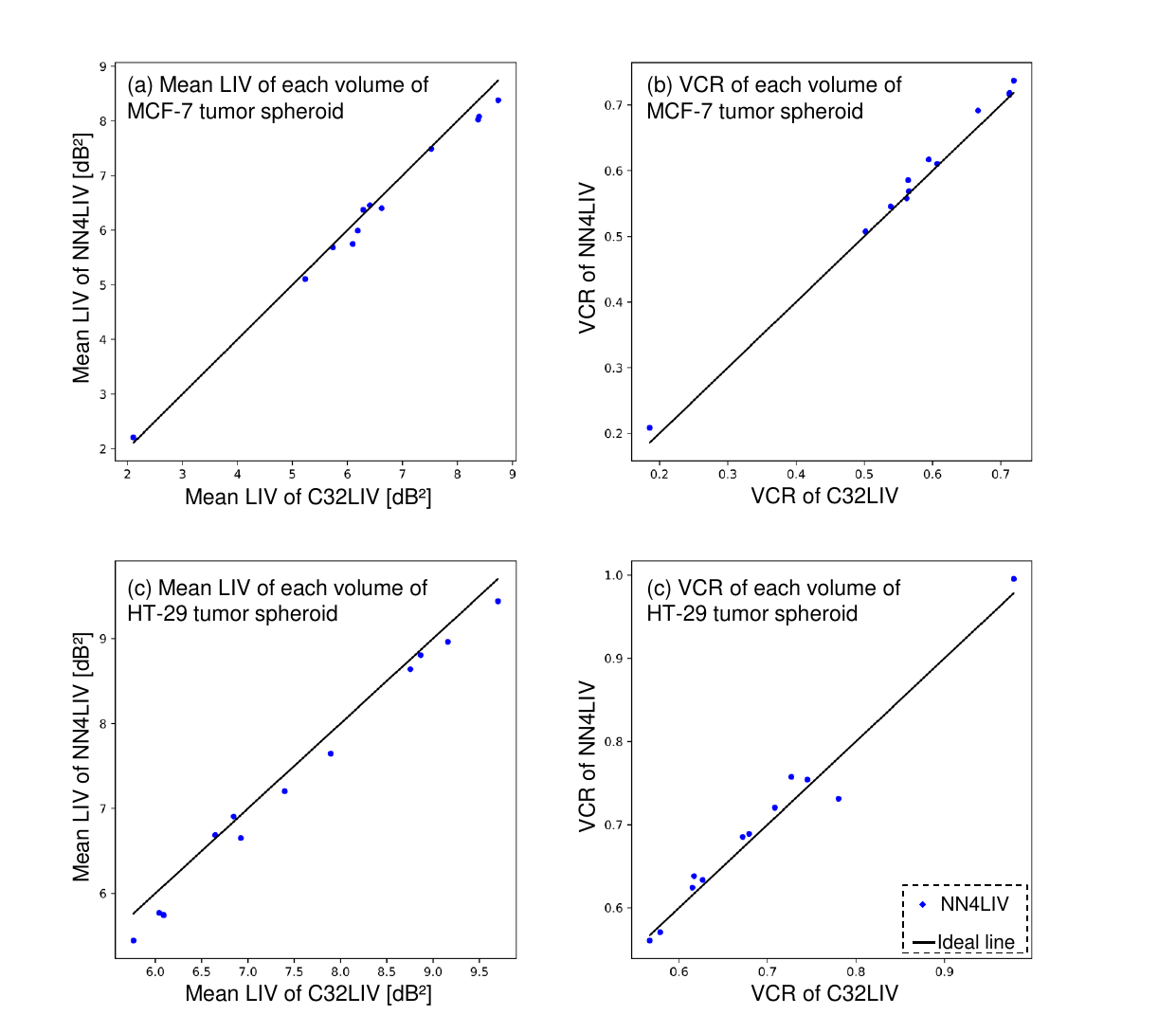}
	\caption{%
		Agreement evaluation of the image matrices obtained from each spheroid type (the first row for MCF-7  and the second row for HT-29). The agreements were evaluated between the metrics computed from NN4LIV (vertical axes) and C32LIV (horizontal axes). The evaluated metrics are the mean volume of each volume (left column) and the VCR or each volume. The metrics are well agreed between NN4LIV and C32LIV.
	}
	\label{Fig:SeparatedMetricsPlots}
\end{figure}
To investigate the sample-type dependency of the NN4LIV, we divided the dataset into two subgroups of MCF-7 and HT-29 spheroids. 
The mean LIV and VCR of the entire spheroid region are plotted in Fig.\@ \ref{Fig:SeparatedMetricsPlots} for each subgroup, and ICC analyses were performed for each group, with the results summarized in Table \ref{tab:SampleDependetAgreement}.
Note that the core- and periphery-ROI analyses were not used to investigate the sample-type dependency, because the necrotic core was formed only with MCF-7.

For both spheroid types, the ICC values for mean LIV and VCR indicated excellent agreement (ICC $>$ 0.9). 
It should be noted that, for HT-29, the mean NN4LIV is slightly negatively shifted from that of C32LIV when C32LIV values are low, as depicted in Fig.\@ \ref{Fig:SeparatedMetricsPlots} (c). 
However, despite this minor negative shift, the ICC remains excellent (ICC = 0.985), and the lower bound of the 95\% confidence interval is still acceptable (0.71). 

\begin{table}[]
	\caption{%
		Spheroid-dependent ICCs and their 95\% confidence intervals (written in [ ]) for the NN4LIV metrics compared to those of C32LIV (the ground truth).
		All ICCs indicate excellent agreement. 
		The lower bounds of the 95\% confidence intervals for the mean LIV of HT-29 spheroids are moderate but still acceptable.
	}
	\centering
	\begin{tabular}{c|c|c}
		\multicolumn{2}{c|}{} & {NN4LIV vs C32LIV}\\
		\hline
		\multirow{2}{*}{MCF-7} & Mean LIV of each volume & 0.992 [0.94, 1.0] \\ \cline {2-3}
		& VCR of each volume & 0.995 [0.92, 1.0] \\ \hline
		\multirow{2} {*} {HT-29} & Mean LIV of each volume &0.985 [0.71, 1.0] \\ \cline {2-3}
		& VCR of each volume & 0.991 [0.91, 1.0]
	\end{tabular}
	\label{tab:SampleDependetAgreement}
\end{table}

\subsubsection{Repeatability of NN4LIV, C4LIV, and C32LIV metrics}
\label{sec:Repeatability}

The repeatabilities of the image metrics of the C32LIV, NN4LIV, and C32LIV images were evaluated by computing the ICCs of the image metrics between the two measurement sessions (S1 and S2).
The resulting ICCs and their 95\% confidence intervals are summarized in Table \ref{tab:repeatability}.

All the C32LIV, NN4LIV, and C4LIV metrics showed ``excellent'' repeatability, i.e., ICC $>$ 0.9.
Note here that the lower bound of the 95\% confidence interval of the mean LIV for C4LIV at the periphery ROI (indicated by red numbers) is only ``good'' (0.84), but this value is still acceptable.
These results indicate that all three types of LIV show high repeatability.
It should also be noted that the agreements (i.e., the ICCs) between the NN4LIV and C32LIV metrics shown in Table \ref{tab:AgreementMain} are close to the repeatability results (ICCs) for C32LIV shown in Table \ref{tab:repeatability}. 
This suggests that the disagreements between the NN4LIV and C32LIV metrics are within the fluctuation range of the C32LIV type itself.

\begin{table}[]
	\caption{%
		ICCs and their 95\% confidence intervals (written in [ ])  for the LIVs from two consecutive measurement sessions (S1 and S2).
		All images show excellent or good repeatability for all image metrics. 
	}
	\centering
	\begin{tabular}{c|c|c|c}
		& \begin{tabular}{c}C32LIV \\(S1 vs S2)\end{tabular} & \begin{tabular}{c}NN4LIV\\ (S1 vs S2)\end{tabular} & \begin{tabular}{c}C4LIV \\(S1 vs S2)\end{tabular}  \\
		\hline
		Mean LIV at core ROI  & 0.998 [0.98, 1.00] & 0.989 [0.90, 1.00] & 0.989 [0.92, 1.00] \\ 
		\hline
		Mean LIV at periphery ROI  & 0.992 [0.94, 1.00] & 0.991 [0.90, 1.00]& 0.977 [\textcolor{red}{0.84}, 1.0] \\ 
		\hline
		Mean LIV of each volume  & 0.994 [0.92, 1.00] & 0.995 [0.96, 1.00]& 0.994 [0.95, 1.00]\\ 
		\hline
		VCR of each volume  & 0.994 [0.97, 1.00] & 0.996 [0.97, 1.00]& 0.995 [0.96, 1.00]\\ 
	\end{tabular}
	\label{tab:repeatability}
\end{table}

\subsection{Demonstration of fast-volumetric measurement}
\label{sec:Demonstration-results}
In all the previous results, we used datasets obtained using the 32-frame-sequence volumetric scan protocol and as a result, the volumetric data acquisition process took 52.4 s, even for the NN4LIV images.
Here, we demonstrate LIV imaging with a volumetric acquisition time of only 6.55 s.
Four untreated beast cancer spheroids that had been cultivated for 11 days were scanned using the fast-volumetric-measurement protocol that was described in Section \ref{sec:Four-frame scanning protocol}.
The NN4LIV and C4LIV images were computed from the datasets obtained via this protocol.
The same trained NN model that was used in the previous sections was used for NN4LIV generation.
In addition, the same spheroids were scanned using the 32-frame-sequence volumetric scan protocol and C32LIV images were then generated for reference.

\begin{figure}
	\centering\includegraphics[width=13cm]{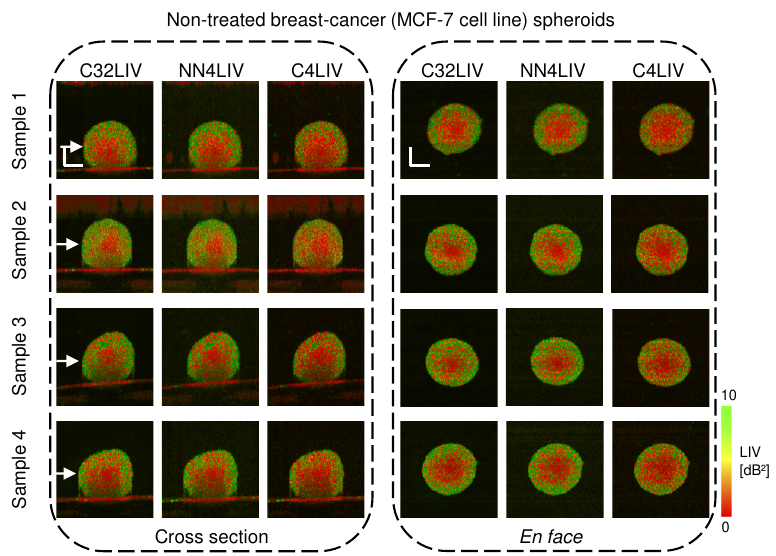}
	\caption{%
		Cross-sectional and \enfaceh NN4LIV and C4LIV images obtained using a 6.55-s volumetric acquisition protocol.
		The samples are four untreated breast cancer (MCF-7) spheroids that were cultivated for 11 days. 
		C32LIV images were obtained from the datasets generated using the 32-frame-sequence volumetric scan protocol, which requires a data acquisition time of 52.4 s/volume.
		The NN4LIV image results reproduce the corresponding C32LIV results well, whereas the C4LIV results show more low-LIV (red) pixels.
		White arrows indicate the depth locations of the \enfaceh images.
		The scale bar represents 200 \um.
	}
	\label{Fig:DemonstrationMCF7}
\end{figure}
The cross-sectional and \enfaceh C32LIV, NN4LIV, and C4LIV images of the four spheroids are shown in Fig.\@ \ref{Fig:DemonstrationMCF7}.
Based on subjective observation, the C32LIV and NN4LIV images show consistent image appearances, whereas the C4LIV images show more low-LIV (red) pixels.

It should also be noted that this demonstration was performed with spheroids after 11 days of cultivation, while the NN model training was performed using spheroids with cultivation times of only 1, 3, and 6 days.
The reasonable generation of the NN4LIV images may suggest generalization of the trained NN model to some degree.

\section{Discussion}
\subsection{NN4LIV enables fast volumetric LIV imaging}
\label{sec:InterpretationOfResults}
It was found that the NN4LIV results resemble the C32LIV (the ground truth) results closely, both qualitatively and quantitatively, as shown in Sections \ref{observationresults} and \ref{AgreementResult}.
In addition, the NN4LIV and C32LIV methods are highly repeatable, as shown in Section \ref{sec:Repeatability}.
Furthermore, as demonstrated in Section \ref{sec:Demonstration-results}, a volumetric NN4LIV tomographic image can be obtained in 6.55 s, while the time required to generate a volumetric C32LIV image is 52.4 s.
Because of its high resemblance to C32LIV, the excellent repeatability, and its compatibility with fast volumetric data acquisition, the NN4LIV method can be substituted for C32LIV and thus enable high-speed volumetric LIV imaging for certain types of tumor spheroids.

In contrast, the C4LIV method was found not to bear such a close resemblance to the C32LIV method, both qualitatively and quantitatively (see Sections \ref{observationresults} and \ref{AgreementResult}).
Therefore, the C4LIV method may not be able to be substituted for C32LIV, although it is compatible with the fast volumetric acquisition protocol.

\subsection{Rationality of the loss function selection}
\label{sec:lossfunctionselection}
\begin{figure}
	\centering\includegraphics[width=13cm]{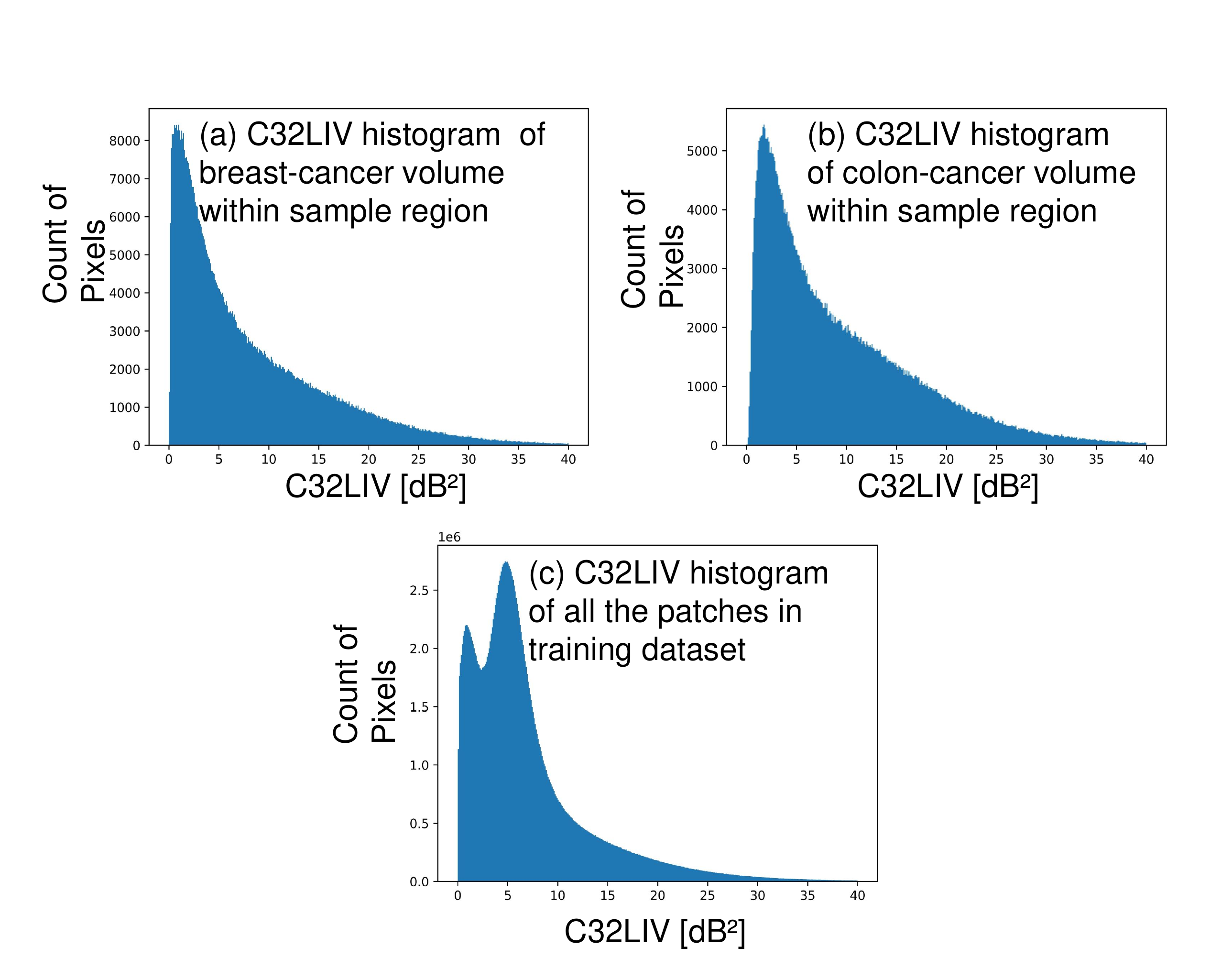}
	\caption{%
		Histograms of C32LIV values (a) in the spheroid region of a breast cancer (MCF-7) volume, (b) in the spheroid region of a colon cancer (HT-29) volume, and (c) of all patches used in the NN-model training process.
		Note that (a) and (b) are the histograms of the spheroid regions only, while the pixels of histogram (c) partially include the nonspheroid region (i.e., the cultivation medium). 
		The left peak in (c) corresponds to the non-spheroid region.
		Highly asymmetric and skewed appearances are shown in all the histograms, and the uneven distribution of the training dataset in (c) hampers effective training of the NN model.
	}
	\label{Fig: Histogram}
\end{figure}
Figure \ref{Fig: Histogram} shows the histograms for (a) the spheroid region of a breast cancer (MCF-7) C32LIV volume, (b) the spheroid region of a colon cancer (HT-29) C32LIV volume (b), and (c) the target C32LIV image patches used for NN model training that partially include the nonspheroid region (i.e., the cultivation medium).
The long-tailed distributions of these histograms indicate that the counts for the high-LIV pixels are far lower than those for the low-LIV pixels.
Note that the small left peak in the target-C32LIV-patch histogram corresponds to the non-spheroid region (i.e., the cultivation medium).
In general, such a highly skewed distribution for the target image (i.e., the ground truth) will hamper effective training of the NN model.

Two well-known strategies can overcome the negative effects of the highly skewed distribution of the ground truth.
The first strategy is re-sampling, and the second is cost-sensitive re-weighting\cite{cui2019class}.
In this study, we used the latter strategy.
In other words, we increased the weights of the high-LIV regions when we computed the losses.
Specifically, we used the wMAE [Eq.\@ (\ref{Eq: wMAE})] as the loss function, as discussed in Section \ref{sec:Training details}, where the wMAE loss is computed by allocating weights of 2 to the pixels with LIV values that are higher than a predefined threshold $\theta$.
Despite the number of high-LIV pixels being relatively small, the allocation of higher weights to the high-LIV pixels may help the NN to focus more strongly on these high-LIV pixels.

To compare the wMAE with other standard loss functions, including the nonweighted mean absolute error (MAE) and the mean squared error (MSE), we trained NN models using these two loss functions.
The training protocol and the datasets used here are the same as those described in Section \ref{sec:implementation}.
In addition, we trained three NN models with the wMAE loss function with predefined thresholds of $\theta$ = 8, 9, or 10 dB$^{2}$, where the 9 dB$^{2}$ threshold is the threshold used in the main study in this paper.
As a result, five NN models were trained in total.
NN4LIV images were generated by all five trained NN models from the S1 datasets of 24 evaluation spheroid datasets.
C32LIV images were also computed from the same evaluation datasets for reference.
The agreements of the image metrics with those of the C32LIV images was quantified using the ICC in a similar manner to that described in Sections \ref{sec:Image evaluation metrics} and \ref{sec:Statistical analysis}.

The ICCs and their 95\% confidence intervals in the cases of MAE, MSE, and wMAE ($\theta$ = 9 \dbsq) are summarized in Table \ref{tab:MAEvsMSE}. 
Although the MAE results show excellent or good agreement levels, the lower bounds of the 95\% confidence intervals are very low (red numbers) for the mean LIVs of both the periphery ROI and of each volume.
The MSE also shows excellent agreement levels, but the lower bounds of the 95\% confidence intervals are low for the mean LIV of the core ROI and the VCR of each volume (red numbers). 
The wMAE with the 9-dB$^2$ threshold, which has been used in the main study, shows excellent agreement, when compared with both the MAE and the MSE, the lower bounds of its 95\% confidence intervals are also very high for all metrics.
We can thus conclude that wMAE is the best loss function among the three types of loss functions under test here.

\begin{table}[]
	\caption{%
		ICCs and their 95\% confidence intervals (written in  [ ])  between the image metrics obtained with the C32LIV images (ground truth) and those obtained with the NN4LIV images generated using the NN models trained using the MAE, the MSE, and the wMAE with the 9-dB$^2$ threshold as loss functions.
		Consideration of the ICCs and the 95\% confidence intervals shows that wMAE provides the best resemblance between the NN4LIV images and the C32LIV images.
	}
	
	\centering
	\begin{tabular}{c|c|c|c}
		& MAE & MSE & \begin{tabular}{c} wMAE\\ ($\theta$ = 9 \dbsq)\end{tabular}\\
		\hline
		Mean LIV at core ROI & 0.914 [\textcolor{red}{0.09}, 0.99] & 0.991 [\textcolor{red}{0.73}, 1.0]  & 0.997 [0.96, 1.0]\\
		\hline
		Mean LIV at periphery ROI & 0.899 [\textcolor{red}{0.01}, 0.99] & 0.990 [0.9, 1.0] &0.988 [0.93, 1.0] \\
		\hline
		Mean LIV of each volume & 0.885 [\textcolor{red}{-0.02}, 0.98] & 0.996 [0.98, 1.0] & 0.991 [0.92,1.0] \\
		\hline
		VCR of each volume & 0.995 [0.98, 1.0] & 0.922 [\textcolor{red}{0}, 0.99] ]& 0.964 [0.96,0.99]\\
	\end{tabular}
	\label{tab:MAEvsMSE}
\end{table}

The NN4LIV image metrics with the different wMAE thresholds were also compared with those of the C32LIV images, with the results as summarized in Table \ref{tab:wMAESelection}.
As shown in the table, the 9-\dbsq wMAE provides excellent agreement for all metrics, and the lower bounds of all the \nnccs with the high threshold are also very high.
However, the lower (8-\dbsq) and higher (10-\dbsq) thresholds provided low lower bounds for the \nnccs for some metrics (red numbers).
As a result, we selected $\theta$ = 9 dB$^2$ for the main study. 
Note that the threshold value was selected based on the specific dataset used in this study.
The optimal threshold may vary among different types of samples, and it may be necessary to re-optimize the wMAE weight to achieve the best performances for other sample types.

\begin{table}[]
	\caption{
		ICCs and their 95\% confidence intervals (written in [ ])  for the image metrics of NN4LIVs generated with wMAE loss functions using the thresholds $\theta$ of 8, 9, and 10 \dbsq versus the image metrics of C32LIV.
		The 9 \dbsq threshold gives excellent ICCs for all image metrics and provides the best lower bounds for the \nnccs among the three thresholds. 
	}
	\centering
	\begin{tabular}{c|c|c|c}
		&\begin{tabular}{c} wMAE\\ ($\theta$ = 8 \dbsq)\end{tabular}& \begin{tabular}{c} wMAE\\($\theta$ = 9 \dbsq)\end{tabular}& \begin{tabular}{c} wMAE\\ ($\theta$ = 10 \dbsq)\end{tabular}\\
		\hline
		Mean LIV at core ROI& 0.998 [0.98, 1.0] & 0.997 [0.96, 1.0] & 0.967 [\textcolor{red}{0.54}, 1.0] \\
		\hline
		Mean LIV at periphery ROI& 0.935 [\textcolor{red}{0.5}, 0.99] & 0.988 [0.93, 1.0] & 0.966 [\textcolor{red}{0.77}, 1.0]\\
		\hline
		Mean LIV of each volume& 0.993 [0.99, 1.0]& 0.991 [0.92,1.0] &0.995 [0.98, 1.0]\\
		\hline
		VCR of each volume & 0.964 [\textcolor{red}{0.05}, 0.99]& 0.964 [0.96,0.99] &0.993 [\textcolor{red}{0.83}, 1.0] \\
	\end{tabular}
	\label{tab:wMAESelection}
\end{table} 

\subsection{Possible improvement of loss function}
\begin{table}[]
	\caption{
		Comparison of the performance between the original wMAE and the modified wMAE, which incorporates OCT intensity information to define weights. Performance is evaluated through ICCs along with their corresponding 95\% confidence intervals (written in [ ]), computed for NN4LIV versus C32LIV (the ground truth). 
		Both variations of wMAE exhibit comparable performance. 
		Notably, the modified wMAE demonstrates a noteworthy improvement of the lower bounds of the 95\% confidence interval for HT-29 mean LIV, which increased from ``moderate'' (0.71) to ``excellent'' (0.92). 
	}
	\centering
	\begin{tabular}{c|c|c|c}
		\multicolumn{2}{c|}{}& Orginal wMAE & Modified wMAE \\ \hline
		\multirow{2}{*}{All spheroids} & Mean LIV of each volume & 0.99 [0.92, 1.0] & 0.992 [0.96, 1.0] \\ \cline{2-4}
		& VCR of each volume & 0.986 [0.96, 0.99] & 0.993 [0.95, 1.0] \\ \hline
		\multirow{4}{*}{MCF-7} & Mean LIV at core-ROI & 0.997 [0.96, 1.0] & 0.983 [0.87, 1.0] \\ \cline{2-4}
		& Mean LIV at periphery-ROI & 0.988 [0.90, 1.0] & 0.985 [0.87, 1.0] \\ \cline{2-4}
		& Mean LIV of each volume & 0.992 [0.94, 1.0] & 0.996 [0.97, 1.0] \\ \cline{2-4}
		& VCR of each volume & 0.995 [0.92, 1.0] & 0.994 [0.9, 1.0] \\ \hline
		\multirow{2}{*}{HT-29} & Mean LIV of each volume & 0.985 [0.92, 1.0] & 0.986 [0.92, 1.0] \\ \cline{2-4}
		& VCR of each volume & 0.991 [0.91, 1.0] & 0.988 [0.94, 1.0] \\ 
	\end{tabular}
	\label{tab:TwoWMAEICC}
\end{table} 

In the present study, we utilized wMAE as a weighted loss function. 
While this approach proved effective, we recognized potential for further enhancement of the loss function. 
One such improvement involves leveraging OCT intensity information to exclude non-sample regions from the learning process, namely, additionally weighting the wMAE to eliminate the contribution of non-sample regions.

Our initial trial of this enhancement involved creating an intensity-based segmentation mask to identify and exclude non-sample (i.e., non-spheroid) regions. 
Specifically, we applied a 13-dB threshold to the OCT intensity image of the first frame within the 32-frame sequence, marking regions below the threshold as non-sample areas. 
Additionally, we manually marked the well plate region as a non-sample area.

A modified version of wMAE was then defined, retaining the original weight definition [Eq.\@ (\ref{Eq: weight})], with the additional specification of setting the weight of non-sample regions to zero.

We trained a neural network model using this modified wMAE and conducted a similar study as described in Section \ref{AgreementResult}, assessing the agreement between NN4LIV and C32LIV. The agreements were evaluated using ICC and summarized in Table \ref{tab:TwoWMAEICC}. The ICCs obtained using the original wMAE were reprinted from Tables \ref{tab:AgreementMain} and \ref{tab:SampleDependetAgreement}, while those obtained using the modified wMAE were obtained using the newly trained model.

As indicated in the table, the modified wMAE demonstrates comparable performance to the original wMAE. 
Although two of the lower bounds of the 95\% confidence interval were not ``excellent,'' they still fall within the ``good'' range. 
Notably, the lower bound of the 95\% confidence interval for HT-29 mean LIV improved from ``moderate'' (0.71) to ``excellent'' (0.92).

Further modification and optimization of weighted loss functions, potentially incorporating more accurate automatic segmentation of sample regions, may lead to further improvements in NN-based LIV generation.

\subsection{Limitations of current NN-based LIV generation method}
There are some limitations and some open issues with the current NN-based DOCT generation method.
One of these limitations is that the current method is specific to a limited range of sample types, i.e., for two types of tumor spheroid.
In addition, the current method can only generate one type of DOCT contrast, i.e., LIV.

For the first issue, the current model was found to be applicable to alveolar organoid samples to some extent, as to be discussed in detail in Section \ref{otherSample}.
However, this validation was performed with only two samples and for a single sample type.
Therefore, more detailed analysis and further generalization are possible future works.

To expand the type of DOCT contrasts, the utilization of more modern NN architectures could be beneficial.
This topic is extensively discussed in Section \ref{discuss:modernNN}.

Another limitation is that it is currently necessary to select the optimal parameters of the wMAE loss function manually via empirical and experimental comparisons, as exemplified in \ref{sec:lossfunctionselection}.
A more universally optimized wMAE loss function and/or a more generalized loss function for a wider variety of samples may be realized through further analysis of the distributions of several datasets.
Additionally, more modern NN architectures, such as generative adversarial networks (GANs) or three-dimensional (3D) convolutional networks, may also help eliminate the need for empirical selection of optimal wMAE parameters as to be discussed in Section \ref{discuss:modernNN}.
This also will be a future work.

\subsection{Applicability to other sample types}
\label{otherSample}
\begin{figure}
	\centering\includegraphics[width=13cm]{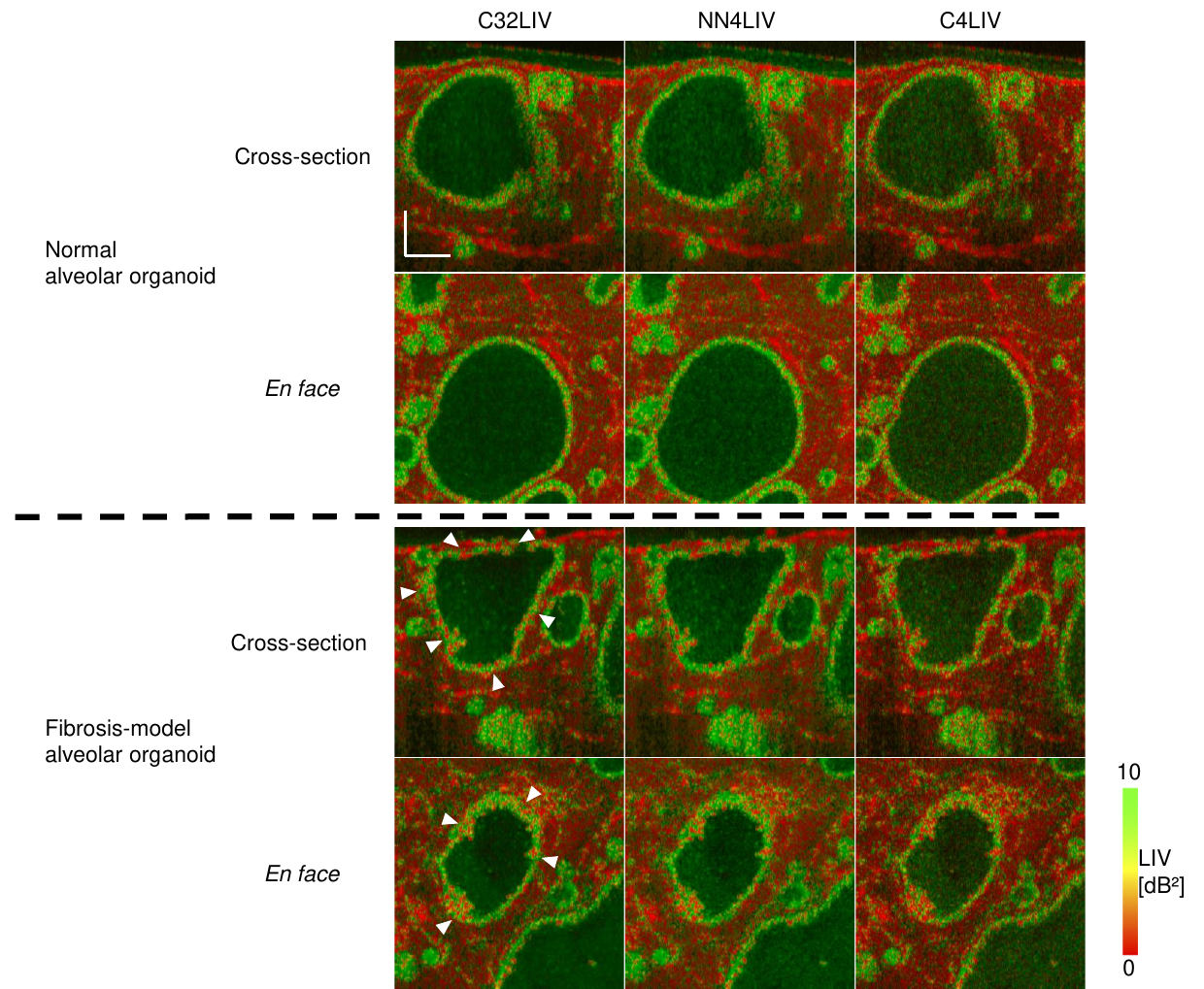}
	\caption{%
		C32LIV, NN4LIV, and C4LIV images obtained from alveolar organoids, including normal organoids (upper half of the figure) and fibrosis model (lower half). 
		These images were generated using the identical NN model employed in the main study. 
		Namely, the model was trained solely on spheroid images. 
		The high resemblance between the NN4LIV and C32LIV indicates that the trained NN model was generalized to another type of sample, namely, alveolar organoids.
		The arrowheads indicate tessellating alveolar epithelia.
		The scale bar indicates 200 \um.
	}
	\label{Fig:AlvelorImages}
\end{figure}
In the main study, the NN model was trained using tumor spheroid images, and evaluation was performed using the same type of samples. 
Here, we investigate the applicability of the trained NN model to other sample types, i.e., other domains.

For this investigation, we used two alveolar organoid samples, which are the same samples used in our previous study\cite{morishita2023label}. 
One is a normal alveolar organoid, which was measured 3 days after its establishment. 
The other sample is a fibrosis-model organoid, also measured 3 days after its establishment but treated with bleomycin for 3 days to induce fibrosis. The detailed cultivation protocol is described elsewhere\cite{yamamoto2017long,morishita2023label}.

The samples were measured using the 1.3-\um JM-OCT system (Section \ref{sec:JMOCT}) with a 32-frame-sequence volumetric scan protocol (Section \ref{sec:32protocol}). 
The lateral FOV is 1 mm $\times$ 1 mm. 
Four OCT frames (8th, 16th, 24th, and 32nd frames) were extracted from the 32-frame sequence, and NN4LIV and C4LIV were computed using the NN model trained with the spheroid data (i.e., the same model used in the main study of this manuscript). 
Note that the NN model has never seen the alveolar organoid samples during its training. 
We also computed C32LIV for reference.

Figure \ref{Fig:AlvelorImages} shows C32LIV, NN4LIV, and C4LIV (first to third columns) of the normal (first and second rows) and the fibrosis-model (third and fourth rows) organoids. 
The first and third rows show cross-sectional images, while the second and fourth rows show \enfaceh images.

Although the structure of the alveolar organoid samples is significantly different from the spheroid samples, namely, the NN model was applied to a significantly different domain, NN4LIV images well recapitulate the C32LIV images. 
On the other hand, C4LIV shows significantly more low LIV (red) pixels than C32LIV and does not closely resemble C32LIV.

The C32LIV of the fibrosis-model organoid shows a tessellating appearance of low and high LIV pixels at the alveolar epithelial regions (arrowheads). This tessellating appearance indicates abnormal remodeling of the epithelium, known as bronchiolization\cite{morishita2023label}. 
It is noteworthy that NN4LIV well recapitulates this abnormal remodeling.

In conclusion, although the NN model was trained for only one domain, i.e., tumor spheroid, it was applicable to another domain, i.e., the alveolar organoid. 
Since this investigation involved only two alveolar organoid samples, we must acknowledge that this conclusion is not strong. 
Future extensive studies may further clarify the applicability of the model to other domains. 
Additionally, including several other types of samples in the training dataset may improve the generalization capability of the NN model.

\subsection{Impact of frame number and inter-frame interval}
\subsubsection{Impact of frame number}
\begin{table}[]
	\caption{%
		Two NN models were trained using two-frame and three-frame sequences, with the corresponding generated LIV denoted as NN2LIV and NN3LIV, respectively. 
		Their agreements with C32LIV (ground truth) were assessed using ICCs and their 95\% confidence intervals (indicated in [ ]). 
		The ICCs of NN4LIV were reprinted from Table \ref{tab:AgreementMain}. 
		While NN2LIV and NN3LIV exhibit good to excellent ICCs, some of the lower bounds of the 95\% confidence intervals are moderate or even poor.
	}
	
	\centering
	\begin{tabular}{c|c|c|c}
		& NN2LIV vs C32LIV & NN3LIV vs C32LIV & NN4LIV vs C32LIV\\
		\hline
		\makecell{Mean LIV at\\ core ROI} & 0.855 [0.27, 0.98] & 0.953 [0.36, 1.0]  & 0.997 [0.96, 1.0]\\
		\hline
		\makecell{Mean LIV at\\ periphery ROI} & 0.926 [0.55, 0.95] & 0.985 [0.86, 1.0] & 0.988 [0.93, 1.0 \\
		\hline
		\makecell{Mean LIV of\\ each volume} & 0.984 [0.96, 0.99] & 0.994 [0.99, 1.0] & 0.991 [0.92, 1.0] \\
		\hline
		\makecell{VCR of\\ each volume} & 0.965 [0.48, 0.99] & 0.984 [0.79, 1.0] & 0.964 [0.96, 0.99] \\
	\end{tabular}
	\label{tab:ImpactOfN}
\end{table}
For investigating the required frame number of our method, we trained two NN models by using two-frame and three-frame OCT sequences. 
The two-frame sequence comprises the first and last OCT frames of the four-frame dataset used in the main study, while the three-frame sequence consists of the first, second, and last frames of the same dataset. 
It is important to note that all frame sequences maintain a consistent time window of 4.92 s.

For evaluation purposes, we fed the model trained with the two-frame datasets with the 8th and 32nd frames from the 32-frame sequence in the evaluation dataset, ensuring that the inter-frame interval matched that of the corresponding training dataset. 
We denote the output of this model as NN2LIV. Similarly, we fed the three-frame trained model with the 8th, 16th, and 32nd frames from the 32-frame sequence, denoting the output as NN3LIV. 
Again, the inter-frame intervals are identical to those of the corresponding training datasets. 
All cases from the evaluation dataset in the main study were included in this additional analysis.

The agreement with the ground truth (C32LIV) was assessed by computing ICCs, with results summarized in Table \ref{tab:ImpactOfN}. 
ICCs were computed for the core- and periphery-ROIs, mean LIV of each volume, and mean VCR of each volume. 
The ICCs of NN4LIV were reprinted from Table \ref{tab:AgreementMain} for reference. 
As discussed in Section \ref{AgreementResult}, NN4LIV exhibited excellent agreements (ICC $>$ 0.9) and very high (excellent) lower bounds of 95\% confidence intervals. 
Both NN2LIV and NN3LIV also demonstrated excellent to good ICCs, but some of the lower bounds of the 95\% confidence intervals were moderate (0.5$ \leq$ ICC $<$ 0.75) or even poor (ICC $<$ 0.5) in some cases. 
Therefore, NN2LIV and NN3LIV are not suitable substitutes for C32LIV.

\subsubsection{Impact of inter-frame time}
\begin{figure}
	\centering\includegraphics[width=13cm]{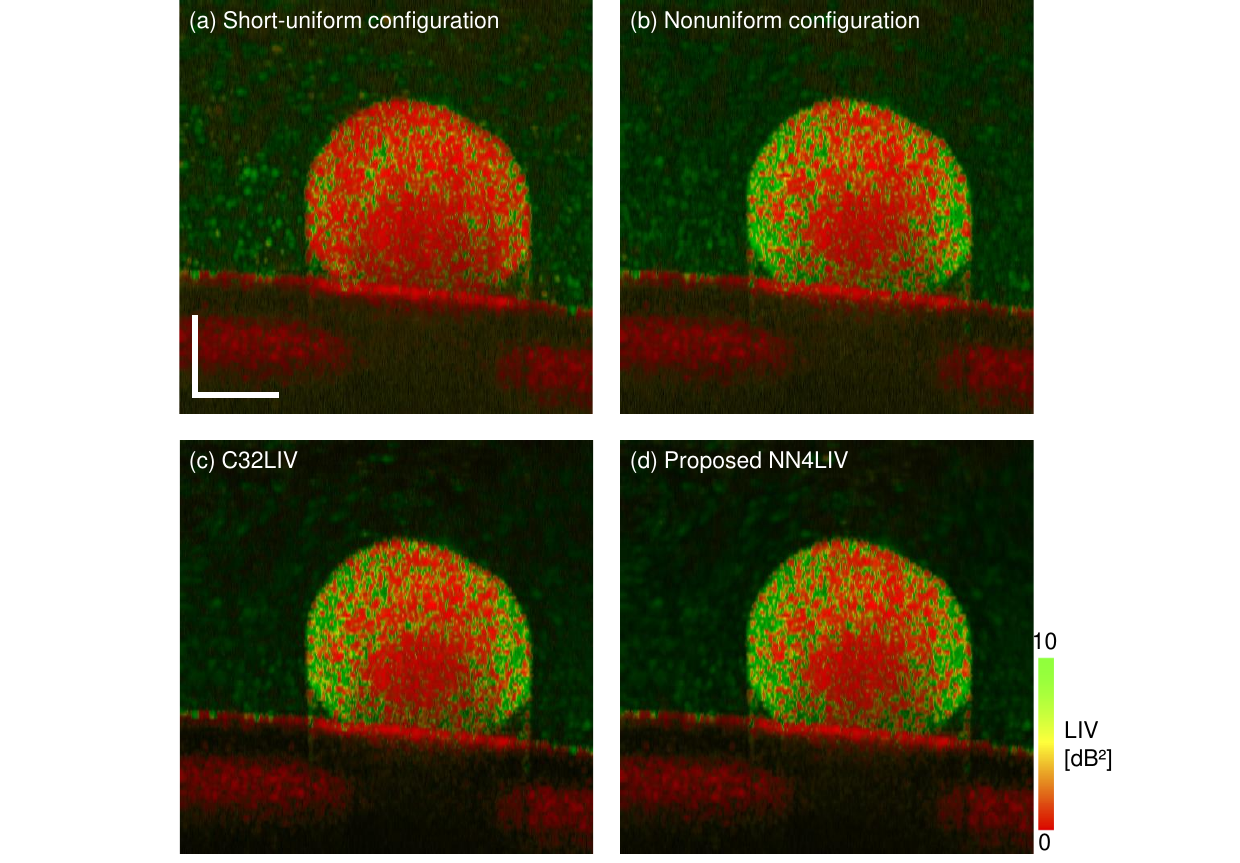}
	\caption{%
		Cross-sectional LIV images generated from a four-frame sequence with short uniform inter-frame intervals (referred to as the short-uniform configuration, panel a) and from a four-frame sequence with short-and-long nonuniform inter-frame intervals (referred to as the nonuniform configuration, panel b) are compared against C32LIV (c) and NN4LIV (d). While the nonuniform configuration yields an acceptable LIV image, the short-uniform configuration fails to produce a meaningful LIV image.
		The scale bar indicates 200 \um.
	}
	\label{Fig:TwoNewConfigNN4LIV}
\end{figure}

\begin{table}[]
	\caption{%
		The ICCs and their corresponding 95\% confidence intervals (indicated in [ ]) comparing the LIV images generated from the short-uniform configuration and the nonuniform configuration versus C32LIV (the ground truth) are presented. 
		The four-frame LIV images were generated using the NN model trained in the main study.
		While short-uniform configuration does not yield acceptable ICCs, those of the nonuniform configuration are excellent. 
		It should be noted that while some lower bounds of the 95\% confidence intervals for the nonuniform configuration are not excellent, they are still considered good.
	}
	
	\centering
	\begin{tabular}{c|c|c}
		& \makecell{Short-uniform\\configuration vs C32LIV} & \makecell{Nonuniform\\configuration vs C32LIV}\\
		\hline
		Mean LIV at core ROI & 0.820 [-0.04, 0.98] &0.986 [0.88, 1.0]  \\
		\hline
		Mean LIV at periphery ROI & 0.12 [-0.02, 0.64] & 0.981 [0.85, 1.0] \\
		\hline
		Mean LIV of each volume & 0.167 [-0.05, 0.52] & 0.985 [0.8, 1.0] \\
		\hline
		VCR of each volume & 0.225 [-0.04, 0.61] & 0.995 [0.99, 1.0] \\
	\end{tabular}
	\label{tab:ImpactOfFrameSeparation}
\end{table}

In the main study, the four frames used to feed the NN were sampled with a uniform inter-frame interval of 1.64 s, resulting in a total time window of 4.92 s. 
Here, we investigated the flexibility of sampling time points and the total time window when feeding the sequence to the original NN model trained with the four-frame sequence featuring the uniform inter-frame interval of 1.64 s.

We employed two new types of four-frame sequences. 
One sequence type consists of the 8th, 9th, 10th, and 11th frames of the original 32-frame sequence, yielding a short inter-frame interval of 204.8 ms and a total time window of 614.4 ms.
We denote this as ``short-uniform configuration.''
The other sequence type, denoted as``nonuniform configuration,'' consists of the 8th, 9th, 31st, and 32nd frames. 
Thus, the inter-frame interval is nonuniform, comprising both a short interval of 204.8 ms and a long interval of 4.5056 s. 
However, the total time window remains identical to that of the original four-frame sequence (i.e., 4.92 s). 

LIV images are generated by feeding these datasets to the NN model used in the main study, which was trained on the original four-frame sequences with a uniform inter-frame interval of 1.64 s. 
The agreement with the ground truth (C32LIV) was assessed using the same evaluation methods as in Section \ref{AgreementResult}.

Figure \ref{Fig:TwoNewConfigNN4LIV} illustrates example LIV images obtained with the short-uniform configuration (a) and nonuniform configuration (b), along with C32LIV (the ground truth) (c) and the standard NN4LIV (d). 
Additionally, Table \ref{tab:ImpactOfFrameSeparation} summarizes the ICCs and corresponding 95\% confidence intervals comparing the two new configurations with C32LIV. 
The image from shot-uniform configuration lacks clear contrast between the necrotic core and viable rim, resulting in an almost poor ICC with C32LIV. Conversely, the image from nonuniform configuration exhibits similar contrast to C32LIV and also to NN4LIV. 
However, some of the lower bounds of the 95\% confidence intervals of ICC in certain metrics were only good rather than excellent.

This study suggests that the trained NN model can generate an LIV image from other types of four-frame sequences. 
However, if the total time window is too short, the generation may fail.

In this study, utilizing the original type of four-frame sequence with a uniform inter-frame interval for both training and inference (i.e., image generation) yielded the best results. 

Nonetheless, frame sequences with nonuniform intervals, such as nonuniform configuration, may possess a wider range of time-frequency components, including higher frequency components than the original four-frame sequences. This potential may be exploited through future optimization of network architecture, training methods, and fine selection of inter-frame intervals.

\subsection{Future directions with modern  NN architectures}
\label{discuss:modernNN}
In the present study, we utilized a well-established but relatively old standard U-net architecture. 
However, in future studies, employing more modern network architectures may enhance the quality of generated images.

One potential improvement involves incorporating a discriminator network, as seen in GANs\cite{goodfellow2020GAN}. 
This approach could obviate the need for wMAE, thereby eliminating the empirical selection of threshold values and weight of wMAE.

Additionally, recurrent neural networks (RNNs)\cite{hochreiter1997LSTM} or 3D convolutions\cite{tran20153Dconv}, which involve temporal convolutions, could effectively capture time-sequential information, which is crucial for certain DOCT methods. For instance, DOCT methods like correlation decay speed\cite{ElSadek2020BOE}, speckle fluctuation spectroscopy\cite{oldenburg2015Optica}, power spectral analysis\cite{apelian2016dynamic,leung2020imaging,munter2020dynamic,Munter2021BOE}, authentic LIV, and swiftness\cite{morishita2024Bios} exploit the sequential order of OCT signals. Leveraging RNNs and 3D convolutions may enable the generation of these DOCT contrasts using neural networks.

Furthermore, employing a multi-input and multi-output neural network architecture, such as a multi-branch GAN\cite{zheng2021MultibrunchGAN}, could enable the integration of multiple input image types (e.g., OCT intensity, phase, and power spectrum). 
Additionally, it could facilitate the simultaneous generation of multiple DOCT contrasts. 
Leveraging multiple input image types may lead to high-quality DOCT images, while simultaneous generation of multiple DOCT contrasts instruct the model to effectively extract temporal and spatial features from the input signals.

\section{Conclusion}
We have demonstrated an NN-based DOCT method that generates LIV images from a small number (typically four) of OCT frames while maintaining similar image quality and similarly high fidelity to conventional LIV images, which are computed from far larger numbers (typically 32) of frames.
Additionally, while the conventional method requires a volumetric acquisition time of 54.2 s/volume, the proposed NN-based method requires only 6.55 s/volume.
Qualitative image comparisons and quantitative image-metric analyses of the NN-based LIV and the conventional LIV confirmed the strong resemblance between these two types of LIV image.
Although the available DOCT contrast types and the measurable sample types are limited at present, the NN-based DOCT method enables high-speed volumetric DOCT imaging with an acquisition time that is eight times shorter than the conventional LIV method.

\section*{Acknowledgments}
The raw OCT data of the alveolar organoids used in Section \ref{otherSample} were originally acquired  for a study published elsewhere\cite{morishita2023label}, and the alveolar organoid samples were provided free of charge by HiLung and Toshio Suzuki.

\section*{Funding}
Core Research for Evolutional Science and Technology (JPMJCR2105); 
Japan Society for the Promotion of Science (21H01836, 22K04962, 22KF0058). 

\section*{Disclosures}
Liu, El-Sadek, Makita, Yasuno: Sky Technology (F), Nikon (F), Kao Corp. (F), Topcon (F), Panasonic (F). 
Mori, Furukawa, Matsusaka: None.

\section* {Data, Materials, and Code Availability} 
Data underlying the results presented in this paper are not publicly available at this time but may be obtained from the authors upon reasonable request. 

\section*{Supplemental document}
See Supplement 1 for supporting content.

\bibliography{reference}

\pagebreak


\title{Supplementary Material}
\setcounter{figure}{0}
\renewcommand\thefigure{S\arabic{figure}}   
\setcounter{table}{0}
\renewcommand\thetable{S\arabic{table}}   

\setcounter{section}{0}
\renewcommand\thesection{S\arabic{section}}  

\begin{abstract}
	In this supplementary material, we present additional cross-sectional and \enfaceh C32LIV, NN4LIV, and C4LIV images.
	LIV images from S2 data of Day-1 breast cancer (MCF-7) spheroids and Day-1 colon cancer (HT-29) spheroids are shown in Fig. \@ \ref{Fig:S2Day1MCF7} and \ref{Fig:S2Day1HT29}, respectively. 
	LIV images from S1 and S2 data of Day-3 breast cancer (MCF-7) spheroids are shown in Fig. \@ \ref{Fig:S1Day3MCF7} and \ref{Fig:S2Day3MCF7}, respectively. 
	LIV images from S1 and S2 data of Day-3 colon cancer (HT-29) spheroids are shown in Fig. \@ \ref{Fig:S1Day3HT29} and \ref{Fig:S2Day3HT29}, respectively. 
	LIV images from S1 and S2 data of Day-6 breast cancer (MCF-7) spheroids are shown in Fig. \@ \ref{Fig:S1Day6MCF7} and \ref{Fig:S2Day6MCF7}, respectively. 
	LIV images from S1 and S2 data of Day-6 colon cancer (HT-29) spheroids are shown in Fig. \@ \ref{Fig:S1Day6HT29} and \ref{Fig:S2Day6HT29}, respectively. 
\end{abstract}


	\begin{figure}[hb]
		\centering\includegraphics[width=13cm]{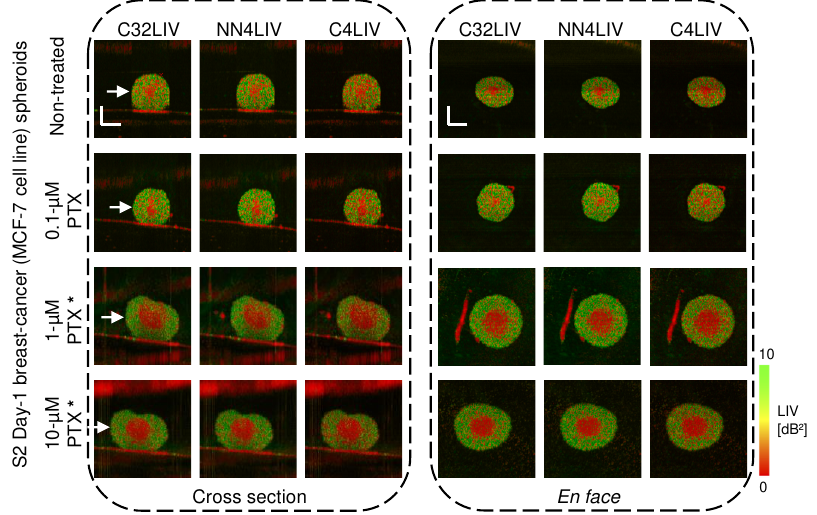} 
		\caption{%
			Demonstration of cross-sectional and \enfaceh images of C32LIV, NN4LIV, and C4LIV of Day-1 breast cancer (MCF-7) spheroids of S2. 
			Cases of both non-treated and PTX treated are shown.
			``*'' indicates the samples used for selecting core-ROIs and periphery-ROIs. 
			White arrow indicates the depth location of \enfaceh images.
			Scale bar indicates 200 \um.}
		\label{Fig:S2Day1MCF7}
	\end{figure}
	
	\begin{figure}
		\centering\includegraphics[width=13cm]{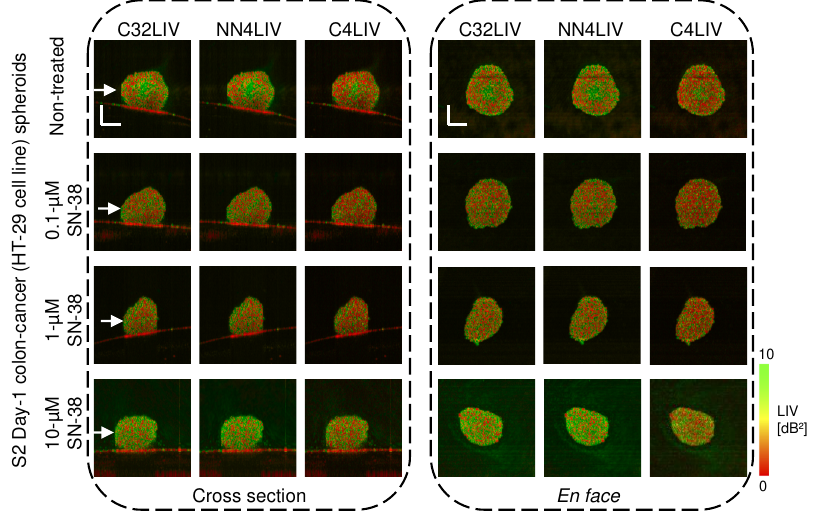} 
		\caption{%
			Demonstration of cross-sectional and \enfaceh images of C32LIV, NN4LIV, and C4LIV of Day-1 colon cancer (HT-29) spheroids of S2. 
			Cases of both non-treated and SN-38 treated are shown.
			White arrow indicates the depth location of \enfaceh images. 
			Scale bar indicates 200 \um.}
		\label{Fig:S2Day1HT29}
	\end{figure}
	
	\begin{figure}
		\centering\includegraphics[width=13cm]{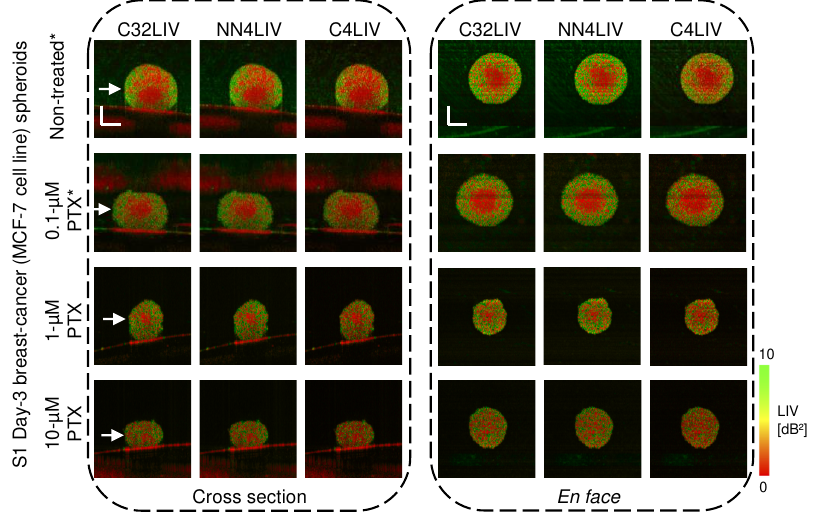} 
		\caption{%
			Demonstration of cross-sectional and \enfaceh images of C32LIV, NN4LIV, and C4LIV of Day-3 breast cancer (MCF-7) spheroids of S1. 
			Cases of both non-treated and PTX treated are shown.
			``*'' indicates the samples used for selecting core-ROIs and periphery-ROIs. 
			White arrow indicates the depth location of \enfaceh images.
			Scale bar indicates 200 \um.}
		\label{Fig:S1Day3MCF7}
	\end{figure}
	
	\begin{figure}
		\centering\includegraphics[width=13cm]{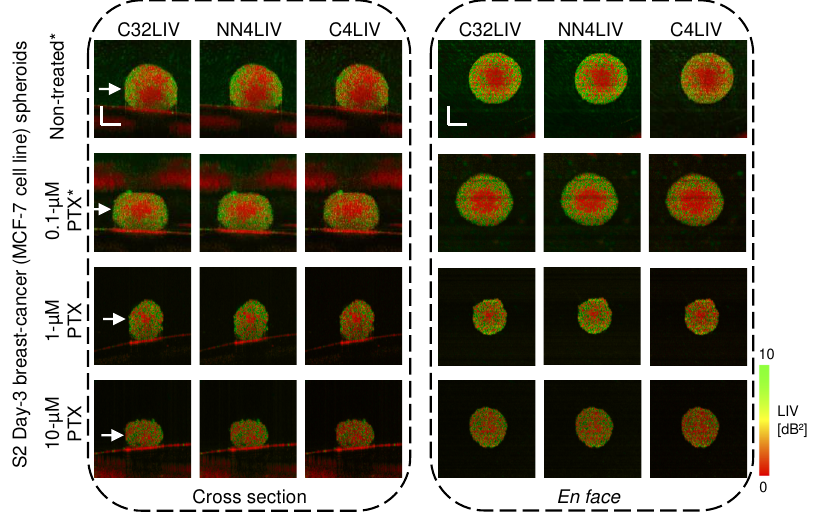} 
		\caption{%
			Demonstration of cross-sectional and \enfaceh images of C32LIV, NN4LIV, and C4LIV of Day-3 breast cancer (MCF-7) spheroids of S1. 
			Cases of both non-treated and PTX treated are shown.
			``*'' indicates the samples used for selecting core-ROIs and periphery-ROIs. 
			White arrow indicates the depth location of \enfaceh images. 
			Scale bar indicates200 \um.}
		\label{Fig:S2Day3MCF7}
	\end{figure}
	
	\begin{figure}
		\centering\includegraphics[width=13cm]{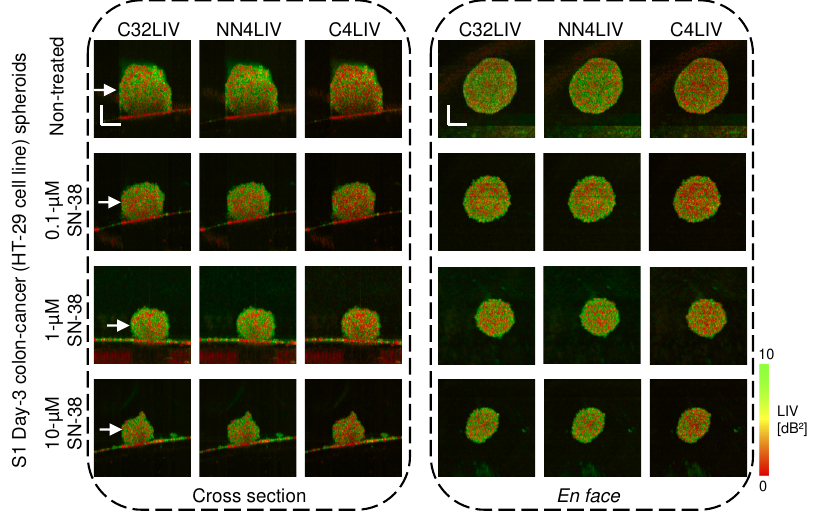} 
		\caption{%
			Demonstration of cross-sectional and \enfaceh images of C32LIV, NN4LIV, and C4LIV of Day-3 colon cancer (HT-29) spheroids of S1. 
			Cases of both non-treated and SN-38 treated are shown.
			White arrow indicates the depth location of \enfaceh images. 
			Scale bar indicates 200 \um.}
		\label{Fig:S1Day3HT29}
	\end{figure}
	
	\begin{figure}
		\centering\includegraphics[width=13cm]{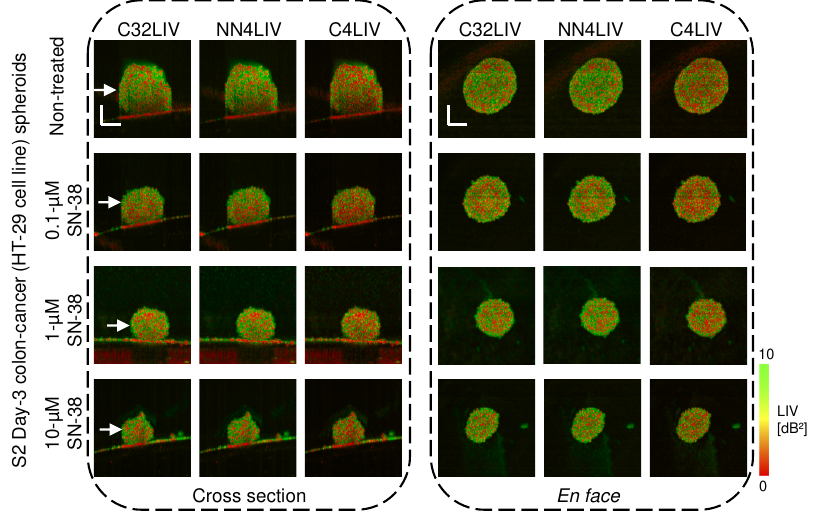} 
		\caption{%
			Demonstration of cross-sectional and \enfaceh images of C32LIV, NN4LIV, and C4LIV of Day-3 colon cancer (HT-29) spheroids of S2. 
			Cases of both non-treated and SN-38 treated are shown.
			White arrow indicates the depth location of \enfaceh images. 
			Scale bar indicates 200 \um.}
		\label{Fig:S2Day3HT29}
	\end{figure}
	
	\begin{figure}
		\centering\includegraphics[width=13cm]{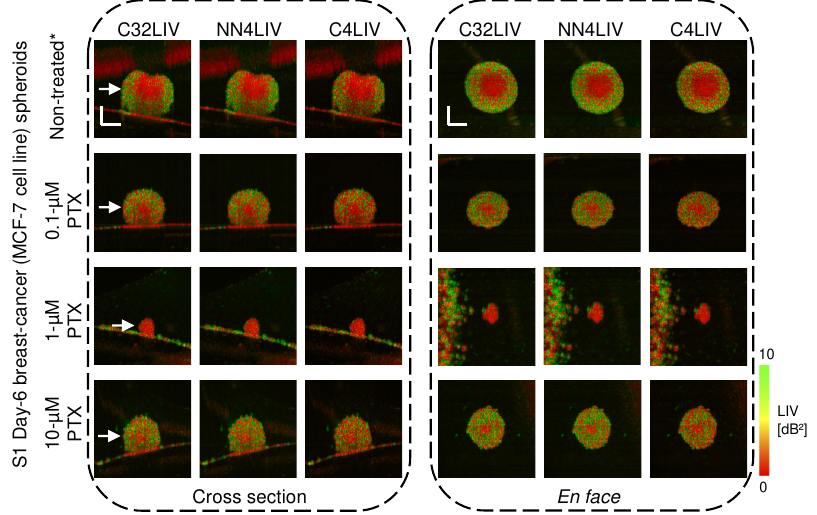} 
		\caption{%
			Demonstration of cross-sectional and \enfaceh images of C32LIV, NN4LIV, and C4LIV of Day-6 breast cancer (MCF-7) spheroids of S1. 
			Cases of both non-treated and PTX treated are shown/
			``*'' indicates the samples used for selecting core-ROIs and periphery-ROIs. 
			White arrow indicates the depth location of \enfaceh images. 
			Scale bar indicates 200 \um.}
		\label{Fig:S1Day6MCF7}
	\end{figure}
	
	\begin{figure}
		\centering\includegraphics[width=13cm]{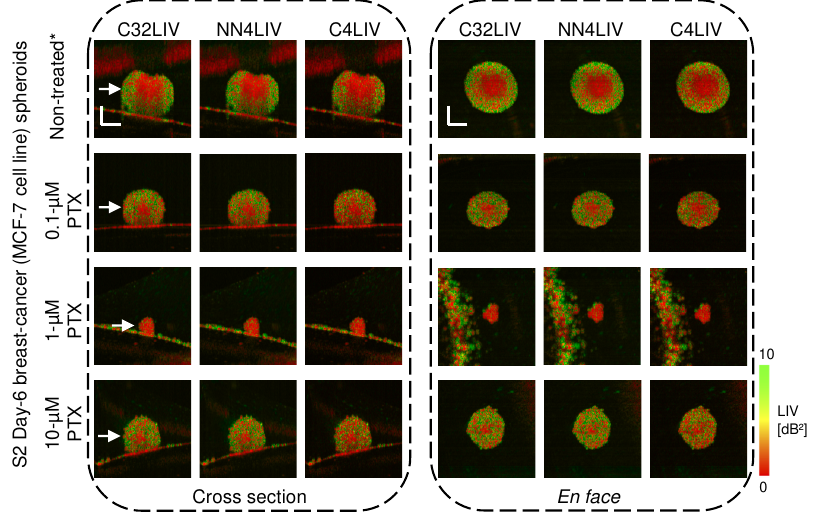} 
		\caption{%
			Demonstration of cross-sectional and \enfaceh images of C32LIV, NN4LIV, and C4LIV of Day-6 breast cancer (MCF-7) spheroids of S2. 
			Cases of both non-treated and PTX treated are shown.
			``*'' indicates the samples used for selecting core-ROIs and periphery-ROIs. 
			White arrow indicates the depth location of \enfaceh images. 
			Scale bar indicates 200 \um.}
		\label{Fig:S2Day6MCF7}
	\end{figure}
	
	\begin{figure}
		\centering\includegraphics[width=13cm]{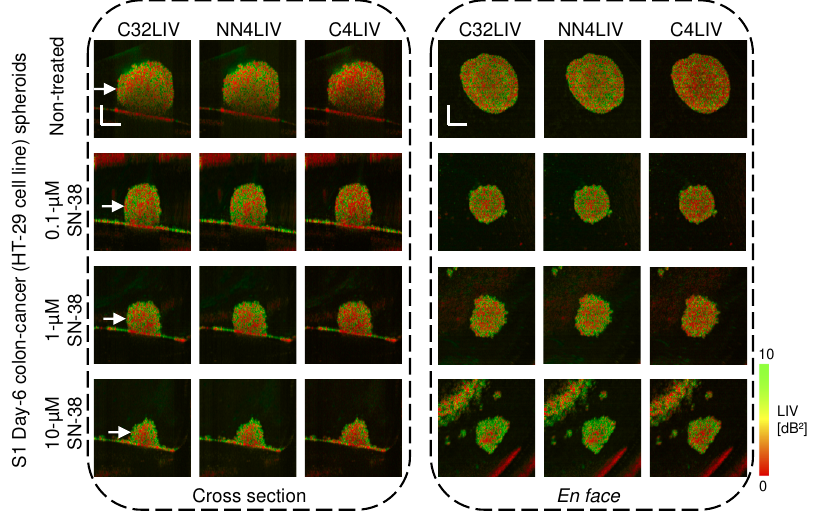} 
		\caption{%
			Demonstration of cross-sectional and \enfaceh images of C32LIV, NN4LIV, and C4LIV of Day-6 colon cancer (HT-29) spheroids of S1. 
			Cases of both non-treated and SN-38 treated are shown.
			White arrow indicates the depth location of \enfaceh images. 
			Scale bar indicates 200 \um.}
		\label{Fig:S1Day6HT29}
	\end{figure}
	
	\begin{figure}
		\centering\includegraphics[width=13cm]{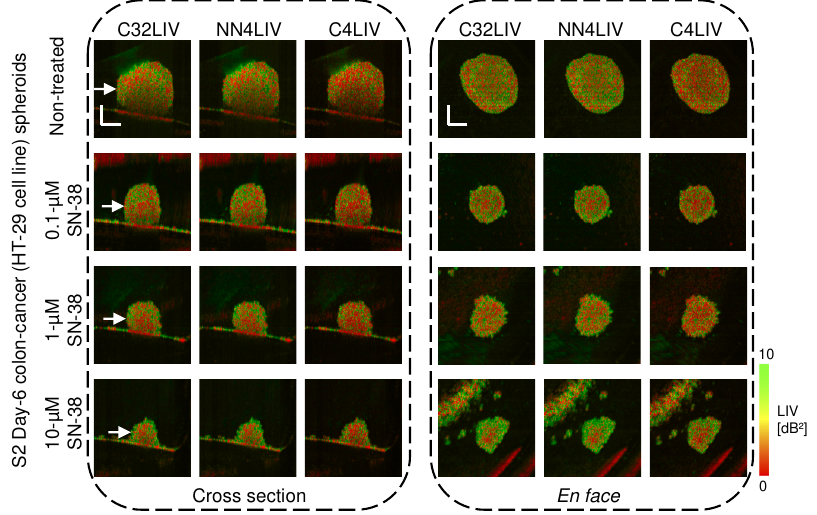} 
		\caption{%
			Demonstration of cross-sectional and \enfaceh images of C32LIV, NN4LIV, and C4LIV of Day-6 colon cancer (HT-29) spheroids of S2. 
			Cases of both non-treated and SN-38 treated are shown.
			White arrow indicates the depth location of \enfaceh images. 
			Scale bar indicates 200 \um.}
		\label{Fig:S2Day6HT29}
	\end{figure}
\end{document}